\documentclass[%
 aip,pof
 amsmath,amssymb,
 reprint,
]{revtex4-2}

\usepackage{graphicx}
\usepackage{dcolumn}
\usepackage{bm}

\usepackage[utf8]{inputenc}
\usepackage[T1]{fontenc}
\usepackage{mathptmx}
\usepackage{etoolbox}
\usepackage{amssymb}
\usepackage{amsmath}
\usepackage{algorithm}
\usepackage{algpseudocode}
\usepackage{hyperref}
\usepackage{xcolor}
\definecolor{color}{HTML}{0000FF}
\hypersetup{linkcolor=color,urlcolor=color,citecolor=color,filecolor=color,colorlinks=true}

\makeatletter
\def\@email#1#2{%
 \endgroup
 \patchcmd{\titleblock@produce}
  {\frontmatter@RRAPformat}
  {\frontmatter@RRAPformat{\produce@RRAP{*#1\href{mailto:#2}{#2}}}\frontmatter@RRAPformat}
  {}{}
}%
\makeatother
\begin{document}

\preprint{AIP/123-QED}

\title[Preprint arXiv:2109.04759]{Monte Carlo simulation of particle size separation in evaporating bi-dispersed colloidal droplets on hydrophilic substrates}
\author{Pavel A.~Zolotarev}
\affiliation{
    Mathematical Modeling Lab., Institute of Physics and Mathematics, Astrakhan State University,  Astrakhan 414056, Russia
}
\author{Konstantin~S.~Kolegov}
 \email{konstantin.kolegov@asu.edu.ru}
\affiliation{
    Mathematical Modeling Lab., Institute of Physics and Mathematics, Astrakhan State University,  Astrakhan 414056, Russia
}
\affiliation{
    Caspian Institute of Maritime and River Transport, Volga State University of Water Transport, Astrakhan 414000,  Russia
}

\date{\today}

\begin{abstract}
Colloidal droplets are used in a variety of practical applications. Some of these applications require particles of different sizes. These include medical diagnostic methods, the creation of photonic crystals, the formation of supraparticles, and the production of membranes for biotechnology. A series of earlier experiments had shown the possibility of particle separation near the contact line, dependent upon their size. A mathematical model has been developed to describe this process. Bi-dispersed colloidal droplets evaporating on a hydrophilic substrate are taken into consideration. A particle monolayer is formed near the periphery of such droplets due to the small value of the contact angle. The shape of the resulting deposit is associated with the coffee ring effect. The model takes into account both particle diffusion and transfers caused by capillary flow due to liquid evaporation. Monte Carlo simulations of such particle dynamics have been performed at several values of particle concentration in the colloidal solution. The numerical results agree with the experimental observations, in which small particles accumulate nearer to the contact line than do the large particles. However, the particles do not actually reach the contact line, but accumulate at a small distance from it. The reason for this is the surface tension acting on the particles in areas where the thickness of the liquid layer is comparable to the particle size. Indeed, the same mechanism affects the observed separation of the small and large particles.
\end{abstract}

\maketitle

\section{Introduction}

The evaporation-induced self-assembly of colloidal particles in droplets and films is a flexible and simple method for obtaining structured deposits. In some applications, solutions containing mixtures of particles of different sizes are of interest. As examples of these, we can include: evaporative lithography~\cite{Harris2009,Utgenannt2013,Kolegov2020,AlMuzaiqer2021}, diagnostics in medicine~\cite{Trantum2011}, the development of biosensors~\cite{Rathaur2020}, the formation of supraparticles~\cite{Liu2019,Gartner2020,Kim2021,Raju2021} and nanocomposites~\cite{Wang2009,Dong2020,Kim2021a}, the creation of photonic crystals~\cite{Choi2010,Liu2019,Nunes2020} and superlattice structures~\cite{Nozawa2022}, the production of color filters for displays~\cite{Das2018}, nanosphere lithography~\cite{Li2009969,Li2009,Chen2009,Utgenannt2016} and inkjet printing~\cite{AlMilaji2019Langmuir}. Mixtures of particles are also used in other technologies for the formation of structures that are not associated with evaporation, for example, the Langmuir-Blodgett method~\cite{Detrich2009,Vogel2011}, spin-coating~\cite{Sharma2009}, the transfer of particle monolayers from a liquid-air interface onto an inclined substrate in the process of liquid discharge from a container with a tap~\cite{Lotito2016}. Particle sedimentation from binary mixtures is also associated with some other interesting phenomena, for example, birefringence, which can be useful in functional materials~\cite{Inoue2020}. Another interesting direction is related to the formation of one-dimensional binary superstructures~\cite{Guo2017}.

Often, the physical properties of the resulting coating are
controlled by adding polymers to the colloidal
solution~\cite{Schulz2020}. Such mixtures allow control of adhesion,
wetting, gloss, biocompatibility, and hydrophilicity. The polymer
concentration and the evaporation rate both affect the properties of
the deposited coating. A mixture of polymers and colloids stratifies
at a relatively high evaporative Peclet number~\cite{Schulz2020}. In
Ref.~\cite{Jeong2021}, this process was simulated by the Brownian
dynamics method. While the evaporation rate can be controlled by the
temperature of the liquid in the film~\cite{Schulz2020}, polymer
particles are sintered at a glass transition temperature below room
temperature~\cite{Samanta2020}. As a result, the colloidal particles
are trapped inside a solid polymeric film after evaporation of the
liquid~\cite{Samanta2020}. Refs.~\cite{Atmuri2012,He2021} include
theoretical and numerical studies of the horizontal stratification
of colloids of different sizes in binary mixtures upon evaporation
of the liquid film. The particle redistribution can be described by
dynamic density functional theory. Continuum equations are
discretized and solved by the finite difference
method~\cite{He2021}. As described in Ref.~\cite{Atmuri2012}
together with theoretical research, an experiment was carried out
with a mixture of colloidal particles of two sizes having different
zeta potentials.  Comparison of two models describing stratification
of the suspension film upon drying has been performed by
Ref.~\cite{Tang2019}. In both models, the motion of the small and
large particles is described using the molecular dynamics method.
The difference between these models is in the explicit and implicit
description of the liquid phase. The calculation results obtained
using both models show that, over time, a layer of small particles
is formed near the free surface of the liquid~\cite{Tang2019}.
Previously, this phenomenon has been explained theoretically based
both on the diffusion of the particles and their interactions. The
range of parameter values at which such stratification is expected
has been determined~\cite{Zhou2017PRL}. A detailed description of
the achievements of the studies of the binary mixture stratification
that occurs during the evaporation of a liquid from a film is given
in the reviews~\cite{Zhou2017,Schulz2018}. For example, the study of
protein stratification during the formation of skin at the
``liquid-air'' interface in a drying bi-dispersed droplet is
potentially valuable for the optimization of milk powder production
in the dairy industry~\cite{Yu2021}.

The dependence of the morphology of micro- and nanoparticle binary
mixture sediments on the concentration of the small particles is
studied in Ref.~\cite{Kumnorkaew2009}. It should be noted that
microparticles form a monolayer in the process of evaporative
self-assembly, but the nanoparticles filling the interstitial spaces
between the microparticles form a multilayered sediment. The effect
of substrate wettability on the separation of the micro- and
nanoparticles in a drying, sessile drop is theoretically and
experimentally studied in Ref.~\cite{Chhasatia2011}. It is possible
to separate by size not only solid `particles', but also liquid
ones. For example, in experiments~\cite{Das2012} bi-dispersed
mixtures of oil droplets inside a sessile water droplet have been
used. The separation of particles with sizes of 1 and 3~$\mu$m was
studied experimentally and computationally in
Ref.~\cite{Devlin2015}. In this experiment, a 2~$\mu$L water drop
was used. Separate placement of the differently sized particles on
the substrate was observed after droplet evaporation. Small
particles were located in the outer sediment ring while the larger
particles accumulated in an inner ring. There was a gap of about
10~$\mu$m between the rings. Each ring was a particle monolayer. The
width of each layer varied from one to several particles across. The
deposition process can be described mathematically using continuum
equations~\cite{Devlin2015}. The authors~\cite{Devlin2015} noted
that the results of their calculations disagree with experimental
observations and therefore concluded that it is necessary to develop
new models. The fact is that their model predicts the rapid growth
of a concentration of large particles near the contact line rather
than the concentration of  small particles here. The distance
between the particles separated by size depends on the contact
angle~\cite{Yi2018}. Ref.\cite{Yi2018} experimented with a droplet
placed on a chemically structured substrate with hydrophilic and
hydrophobic regions. This allowed control of the contact angle
$\theta$ to obtain a smooth contact line. In the
experiment~\cite{Yi2018}, large (1~$\mu$m), medium (500~nm) and
small (100~nm) polystyrene particles were used. Concentric rings of
different particle sizes were observed in the resulting deposits. In
Ref.~\cite{Yi2018}, a ring with small particles was located closer
to the contact line. Furthermore, toward the center of the drop,
there was a ring with medium-sized particles and then a ring with
large particles did. According to the results~\cite{Yi2018} , the
distance between the outer rings was about 8.6, 10.5, and
16.6~$\mu$m for $\theta =$ 50$^\circ$, 30$^\circ$ and 14$^\circ$,
respectively. A series of experiments with droplets on hydrophobic
and hydrophilic substrates is described in Ref.~\cite{Singh2011}.
The authors studied the effect of the size ratio of large and small
particles on the resulting  structure. This study showed the
possibility of creating multilayer crystalline structures from mixed
particles~\cite{Singh2011}. In another study, the goal was to obtain
a monolayer of a particle mixture  for use as a lithographic mask in
structuring deposits formed of gold nanoparticles or of
biomolecules~\cite{Singh2011acsNano}. For example, the self-assembly
of a binary particle mixture can be used to structure proteins on a
surface~\cite{Singh2011advMater}. The method enables the structuring
of relatively large areas~--- up to several square centimeters. This
can be useful in some biological and medical
applications~\cite{Singh2011advMater}. Another possible application
is associated with plasma polymerization of the substrate surface
through a deposited particle monolayer, acting as a mask. Such
processing allows coatings to be obtained with periodic chemical
properties~\cite{Singh2011adfm}. The multilayer structures of binary
and ternary particle mixtures of different sizes were created layer
by layer using evaporative self-assembly~\cite{Singh2011adfm21}. The
formation of supraparticles during the process of bi-dispersed
colloidal droplet drying on a superamphiphobic surface has been
studied in Ref.~\cite{Liu2019}. On such substrates, the contact
angle exceeds 150$^\circ$. Therefore, the droplet shape is close to
spherical. The results of such experiments and molecular dynamics
simulations have shown particle separation occurs during
evaporation. Small particles form an outer layer with a close-packed
crystal structure. The number of large particles increases, and the
inner structure becomes amorphous, toward the center of the formed
cluster. The morphology of such supraparticles is similar to a
core-shell arrangement~\cite{Liu2019}.

\begin{figure}
    \includegraphics[width=0.95\linewidth]{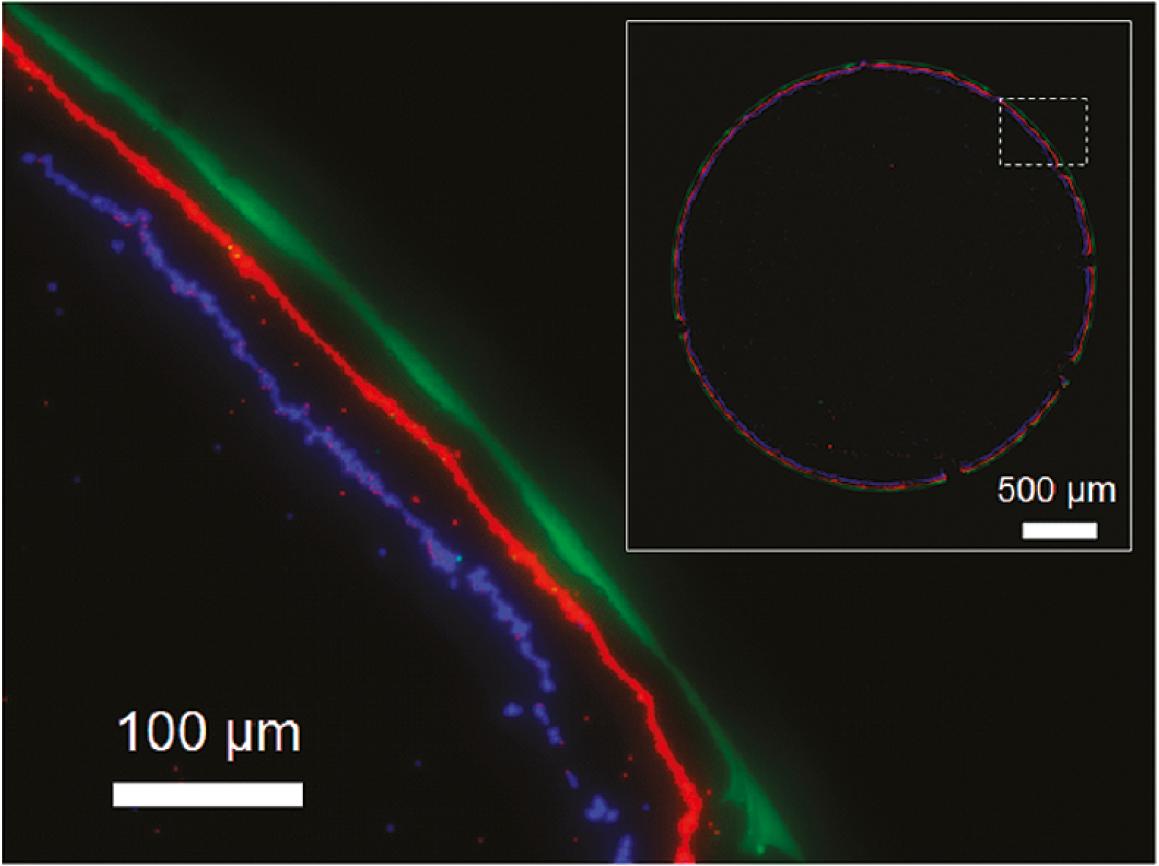}\\
    \caption{
Optical fluorescence image showing the separation of 40~nm (green), 1 $\mu$m (red), and 2 $\mu$m (blue) particles after evaporation. Reprinted with permission from Ref.~\cite{Wong2011} (\copyright 2011, ACS).
}
    \label{fig:WongExperiment}
\end{figure}

The type of sediment remaining after a water droplet has dried also depends on the temperature of the substrate~\cite{Parsa2017}. When a mixture of 1~$\mu$m and 3.2~$\mu$m particles was used in the experiment~\cite{Parsa2017} particle size separation could be observed near the contact line. The authors~\cite{Parsa2017} varied the temperature of the silicon substrate in the range from 22 to 99 $^\circ$C, thus obtaining various deposit structures, from uniform spots to concentric rings. In another study, the temperature of the silicon substrate was varied from 27 to 90 $^\circ$C~\cite{Patil2018}. The values of the contact angles of the water droplets were varied in the range from 3.6 to 65.2$^\circ$ depending on the substrate temperature. Particle sizes in a binary mixture were considered in the range from 100~nm to 3~$\mu$m~\cite{Patil2018}. Depending on the substrate temperature, either Marangoni flows or capillary flows prevailed, and this determined the type of final deposit.  The authors~\cite{Patil2018}   demonstrated the possibility of particle self-sorting near the contact line. The possibility of particle separation in a ternary mixture based on the coffee ring effect (Fig.~\ref{fig:WongExperiment}) has been shown in Ref.~\cite{Wong2011}. Furthermore, these authors carried out the experiment in the context of the separation of biological components. This approach is promising for use in medical diagnostics in the future. Another experiment with an evaporating water drop on glass at $\theta\approx$ 10--15$^\circ$ has been described by Ref.~\cite{Monteux2011}. Small (40 and 100~nm) and large (1, 3 and 5~$\mu$m) particles were transferred to the periphery of the droplet by capillary flow, leading to the formation of an annular deposit. There was a depleted zone between the edge of the deposit and the contact line. Geometric considerations indicated that the width of this zone was determined as $\lambda\approx d_p/ \tan \theta$, which is consistent with experimental measurements~\cite{Monteux2011} ($d_p$ is the particle diameter). Three subregions could be distinguished in the sediment: 1) an accumulation of small particles close to the contact line, then 2) a region of a mixture of different sizes of particle is located, outside which was 3) a zone of large particles. The particle separation was influenced by the contact angle $\theta$ and the particle size~\cite{Monteux2011}. The authors~\cite{Iqbal2018} used both hydrophilic ($\theta\approx 27^\circ$) and hydrophobic ($\theta\approx 110^\circ$) substrates in their experiments. In addition, various ratios of large and small particle sizes were considered. In the case of a hydrophobic substrate, the deposition form appears as a central spot. By contrast, two types of sediment were obtained on a hydrophilic substrate, depending on the particle size~\cite{Iqbal2018}. Annular sediments were observed for a mixture of 0.2 and 3~$\mu$m particles. The particles were separated by size near the contact line, the small particles being located in the outer ring, while the larger ones were concentrated in the inner ring. A small separation was noticeable between the rings. In the case of a 1 and 6~$\mu$m particle mixture, the annular deposition of small particles  was observed near the periphery, while larger particles formed clusters in the inner region of the deposit. The authors have made a theoretical assessment of various types of particle interaction  to explain this phenomenon~\cite{Iqbal2018}. Their hypothesis is based on the ratio of the surface tension and friction forces for the large and small particles.

The effect of thermocapillary flow on the dynamics of large and small particles in an almost spherical droplet on a superhydrophobic surface ($\theta \geqslant 150^\circ$) has  been described in Ref.\cite{Marinaro2021}, where the flow velocity field was measured experimentally. The results of numerical calculations have shown  that large particles move toward the central region, while small particles are transferred mainly out to the periphery of the evaporating droplet. The hydrodynamics has been modeled on the basis of simplified Navier--Stokes equations while the particle dynamics have been described using the Langevin equation~\cite{Marinaro2021}. The possibility of particle sorting during the evaporation of liquid from a capillary bridge formed between two parallel plates has been studied in Ref.~\cite{Upadhyay2021}. Here, hydrophilic and hydrophobic plates were used in different combinations. A stick-slip motion of the contact line leads to the formation of concentric rings. The influence of the solution concentration and different particle sizes was also studied in this work~\cite{Upadhyay2021}. Sedimentation of particles under gravity in a drying sessile/ pendant droplet of a bidisperse suspension on a hydrophobic substrate can affect the suppression of the coffee ring effect, which is extremely important for inkjet printing ~\cite{Hu2021}.

The purpose of our current study is to verify the theoretical explanation of particle separation, as a result of their size, when they are near the contact line in an evaporating droplet on a hydrophilic substrate. An explanation has been suggested in the experimental studies~\cite{Monteux2011,Wong2011,Yi2018} that have shown this phenomenon. The idea is that a particle driven by the capillary flow is not able to reach the contact line. It stops at a short distance in front of the contact line. This distance is determined by the value of the angle $\theta$ and the particle size itself. The boundary that particles cannot cross has been called the fixation radius~\cite{Kolegov2019}. This boundary corresponds to the radial coordinate, where the thickness of the liquid layer is about the same as the particle size. The particle cannot cross the fixation radius, since the surface tension force of the liquid restrains it. We decided that it was necessary to carry out computational experiments to confirm this particle separation mechanism. In this study, numerical calculations are performed using a mathematical model proposed by Refs.~\cite{Kolegov2019,Zolotarev2021}. Here, this model is adapted to a binary particle mixture.


\section{Methods}

\subsection{Physical statement of the problem}

\begin{figure}[!htb]
    \includegraphics[width=0.95\linewidth]{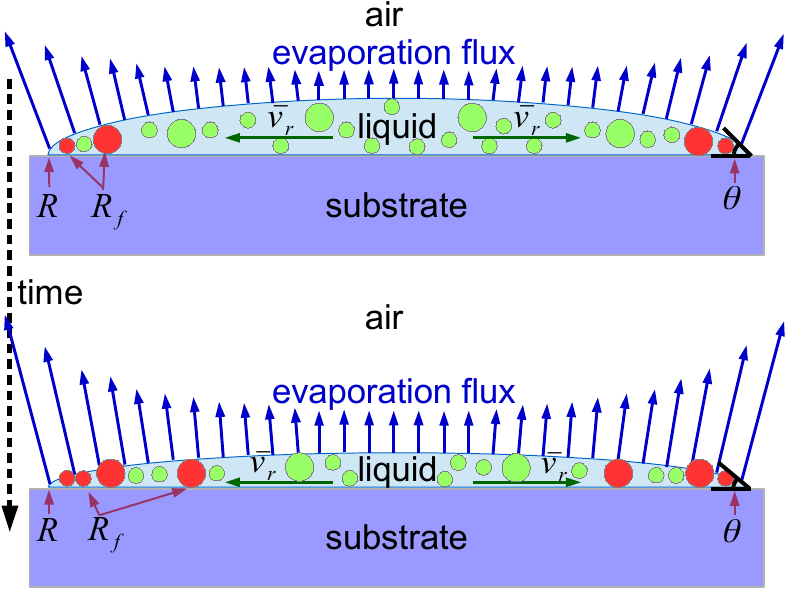}
    \caption{\label{fig:SketchBi-dispersedProblem}
        Sketch to the problem statement. Adapted with permission from Ref.~\cite{Kolegov2019} (\copyright 2019, APS).}
\end{figure}

Consider a colloidal droplet with polystyrene microspheres of two sizes placed on a hydrophilic substrate (Fig.~\ref{fig:SketchBi-dispersedProblem}). Let us denote the radius of the large particles as $r_l$ and the radius of small particles as $r_s$ ($r_s < r_l$). In our assumption, the three-phase boundary is pinned throughout the entire evaporation process. Therefore, the contact  radius of the droplet with the substrate, $R$, is constant. This condition is fulfilled in the case of a rough substrate or a sufficient particle concentration. The glass substrate is impermeable and perpendicular to the direction of the gravity vector.

We do not take into account the effect of gravity on the droplet shape, since the Bond number $\mathrm{Bo}\approx g h_0^2 \rho_l / \sigma \approx 1.4 \times 10^{-3} \ll 1$, where $g$ is the acceleration due to gravity, $h_0\approx R\, \theta / 2$ is the initial droplet height, $\rho_l$ is the liquid density, and $\sigma$ is the surface tension. The problem parameters and their values are given in the Table~\ref{tab:ParametersInBiDispersedDroplet}. The values of the physical parameters for the liquid are taken for water. The time of  droplet evaporation is calculated as $t_\mathrm{max}=\rho_l V/ \dot m$, where $V\approx h_0 R^2$ is the liquid volume. The evaporation rate has been calculated as $\dot m\approx \pi R D_v (1-H) \rho_v (0.27 \theta^2 + 1.3)$, where $D_v$ is the vapor diffusion coefficient, $H$ is the relative vapor pressure (or the humidity in the case of water) and $\rho_v$ is the saturated vapor concentration at the surface temperature of the droplet~\cite{Larson2014}.  The particle diffusion coefficient has been calculated using the Einstein formula $D_{s,l}=kT/(6\pi \eta r_{s,l})$ under the assumption of a weak solution (Table~\ref{tab:ParametersInBiDispersedDroplet}), where $k$ is the Boltzmann constant, $T$ is the liquid temperature, and $\eta$ is the viscosity. Let us estimate the Stokes velocity for large particles, $v_\mathrm{sed}=2 r_l^2 \Delta \rho g / (9\eta)\approx 10^{-7}$ m/s, where $\Delta \rho = \rho_p - \rho_l$ ($\rho_p$ is the density of the polystyrene spheres). Then the sedimentation time of the large particles is $t_\mathrm{sed}= h_0 / v_\mathrm{sed} \approx 10^3$ s. We do not take into account the sedimentation of particles in this problem, since $t_\mathrm{sed} \gg t_\mathrm{max}$, where $t_\mathrm{max}$ is the evaporation time. A theoretical estimation shows that capillary flow prevails over Marangoni flow for small values of $\theta$~\cite{Kolegov2019}. This is also evidenced by experimental work~\cite{Monteux2011,Wong2011,Yi2018}. We do not take into account the thermocapillary flow here for this reason. Now let us estimate the time of diffusion ordering of the small particles, $t_d=d_s^2/D_s\approx$ 2~s, where $d_s=2 r_s$ is the diameter of the small particles. This time is $t_d=d_l^2/D_l\approx$ 17~s for large particles with a diameter $d_l=2 r_l$.  Here, we do take into account the diffusion of the particles, since $t_d \ll t_\mathrm{max}$ in both cases. In addition, within the framework of the 2D model, taking diffusion into account will partially compensate for the lack of particle freedom that would exist in a 3D space. Let us also estimate the value of the Stokes number $\mathrm{Stk}=\rho_p d_{s,l}^2 v_c/ (18 \eta L_c)$, where $v_c$ is the characteristic flow velocity and $L_c$ is the characteristic distance. As a rule, the velocity $v_c$ varies in the range of 1--10 $\mu$m/s in the case of a water drop under room conditions.  We will consider $L_c$ as the radius of the droplet, $L_c=R$. Thus, the largest value of the Stokes number is  $\mathrm{Stk}\approx 10^{-9}$. It allows us to conclude that the particle velocity and the velocity of the fluid flow coincide, since $\mathrm{Stk}\ll 1$. Here, we do not take into account the various possible types of ``particle--particle'' and ``particle--substrate'' interactions (capillary, electrostatic, and molecular interactions)~\cite{Li2021}, since the theoretical explanation of particle separation by size near the contact line is not associated with these effects in Refs.~\cite{Monteux2011,Wong2011,Yi2018}. In addition, taking into account these effects would greatly complicate the model. Here, we only pretend to a phenomenological explanation of the particle separation effect. In this paper, we consider the case when the deposit is a monolayer of particles. The parameter values have been chosen from Ref.~\cite{Wong2011} for our simulation since a sufficiently small value of the contact angle ($\theta\approx 9.5^\circ$), at which the monolayer sediment is formed, was observed in that study. In Ref.~\cite{Wong2011}, three sizes of particle were used: 40~nm, 1~$\mu$m, and 2~$\mu$m in diameter (Fig.~\ref{fig:WongExperiment}). However, the principle of particle separation in binary and ternary solutions is identical. Therefore, we can consider a binary particle mixture without loss of general applicability, although, it is necessary to use a 3D model to take into account the case with the ratio $\gamma = r_s / r_l \ll 1$, because small particles can pass through the pores existing between closely spaced large particles. This, further, applies to the case with multilayered particle sediments, in which it is also necessary to take into account the vertical motion of particles.
\begin{table}
\caption{Parameters of the problem.}
\centering
\begin{tabular}{|p{0.13\linewidth}|p{0.45\linewidth}|p{0.36\linewidth}|}
  \hline
  Symbol& Parameter& Value/ Unit of measure\\
  \hline
  $\eta$& Liquid viscosity& $0.9 \times 10^{-3}$ [s~Pa]\\
  $\sigma$& Surface tension& $73 \times 10^{-3}$ [N/m]\\
  $g$& Acceleration due to gravity& 9.8 [m/s$^2$]\\
  $\rho_l$& Liquid density& $1000$ [kg/m$^3$]\\
  $\rho_p$& Particle density& $1050$ [kg/m$^3$]\\
  $T$& Liquid temperature& 295 [K]\\
  $r_s$& Small particle radius& 0.5 [$\mu$m]\\
  $r_l$& Large particle radius& 1 [$\mu$m]\\
  $R$& Droplet radius& 1.375 [mm]\\
  $\theta_0$& Contact angle& $\pi / 18.95$ [rad]\\
  $h_0$& Initial droplet height& 0.1 [mm]\\
  $t_\mathrm{max}$& Evaporation time& 157 [s]\\
  $k$& Boltzmann constant& $1.38\times 10^{-23}$ [J/K]\\
  $D_s$& Diffusion coefficient for small particles& $4.8\times 10^{-13}$ [m$^2$/s]\\
  $D_l$& Diffusion coefficient for large particles& $2.4\times 10^{-13}$ [m$^2$/s]\\
  $H$& Relative humidity& 0.46\\
  $\rho_v$& Saturated vapour concentration& $18.0\times 10^{-3}$ [kg/m$^3$]\\
  $D_v$& Vapor diffusion coefficient& $2.2\times 10^{-5}$ [m$^2$/s]\\
  \hline
\end{tabular}
\label{tab:ParametersInBiDispersedDroplet}
\end{table}

\subsection{Model description}
The model~\cite{Kolegov2019,Zolotarev2021} is semi-discrete since hydrodynamics is modeled within a continuum approach, while the motion of each colloidal particle is explicitly considered. Here, we adapt this model for the case of two sizes of particles. The geometric region under consideration is depicted in a circle with a radius of $R$ (view of the droplet from the top). The real 3D region is close to a spherical segment in shape for the case, where capillary force predominates over gravity. Consider a thin layer of liquid ($h_0 \ll R$), as we pass from the 3D formulation of the problem (spherical segment) to the 2D formulation (circle) where the particles can only  move in the horizontal plane. Therefore, their positions are given as $(x,y)$ in the Cartesian coordinate system, or as $(r,\varphi)$ in a polar coordinate system. Point $(x=0, y=0)$ corresponds to the center of the circle. Small and large particles in the 2D formulation are described as circles with radii of $r_s$ and $r_l$, respectively, inside a large circle with radius of $R$ ($r_s < r_l \ll R$). From the approximation of the droplet shape, $$h(r,t) \approx \theta (t) \frac{R^2 - r^2}{2R},$$ we express the fixation radius $$R_{s,l} (t) \approx \sqrt{R^2 - \frac{4 r_{s,l} R}{\theta (t)}},$$ where $\theta(t) \approx \theta_0 \left( 1 - t/ t_\mathrm{max} \right)$. It should be noted that $R_l(t) < R_s(t) < R$. We assume that a particle cannot cross its fixation radius in the process of motion since this is counteracted by the capillary force. On the other hand, we assume that the boundary $R_{s,l}$ can cross the particle without dragging it along. Otherwise, this would lead to the formation of a central spot deposition, which is not observed in the experiment~\cite{Wong2011}.

Here, particle diffusion is simulated by the Monte Carlo method. A random polar angle $\alpha \in [-\pi; \pi)$ is generated at each time step $\tau$ and for each particle. The new position of the particle is calculated using  the formulas $x_{\tau +1}^{s,l} = x_\tau^{s,l} + \cos (\alpha)\sqrt{2 D_{s,l}\, \delta t}$ and $y_{\tau +1}^{s,l} = y_\tau^{s,l}  + \sin (\alpha)\sqrt{2 D_{s,l}\, \delta t}$. In Refs.~\cite{Kolegov2019,Zolotarev2021}, the time step value, $\delta t =$ 0.1~ms, was selected on the basis of a series of computational experiments so that the Einstein relation for the mean square displacement of particles~\cite{Ortega2010,Deshmukh2015} with a radius of 0.35~$\mu$m was satisfied. In the present study, the particle size is approximately 1.5--3 times larger. Therefore, the time step $\delta t =$ 0.1~ms is sufficiently small to approximate the Brownian motion of the particles.

Now we shall consider the transfer of particles caused by the capillary flow of liquid. Since $\mathrm{Stk}\ll 1$, the particle velocity is equal to the fluid flow velocity calculated using the approximate analytical formula,
$$
\bar v_r (r,t)=
\frac{R}{4 \tilde r (t_\mathrm{max}-t)}\left[  \frac{1}{\sqrt{1-\tilde r^2}} - \left( 1-\tilde r^2 \right) \right],
$$
obtained from the mass conservation law (a detailed derivation of the formula and references to primary sources are given in Ref.~\cite{Kolegov2019}). Here, $\tilde r = r/R$ is used, and $\bar v_r$ is the velocity of the radial fluid flow averaged over the droplet height.  As a result of drift of the particles caused by the fluid flow, the radial coordinate of a particle changes at the next time step,
\begin{equation*}
    r_{\tau+1}^{s,l} =
    \begin{cases}
        r_\tau^{s,l} + \bar v_r \delta t, \; r_\tau^{s,l} + \bar v_r \delta t\leq R_{s,l} \\
        R_{s,l},\; r_\tau^{s,l} + \bar v_r \delta t > R_{s,l}.
    \end{cases}
\end{equation*}
The condition $r_{\tau+1}^{s,l} > R_{s,l}$ prevents the particle from crossing the fixation radius.

\subsection{Problem-solving algorithm}
\begin{figure}[!htb]
    \includegraphics[width=0.49\linewidth]{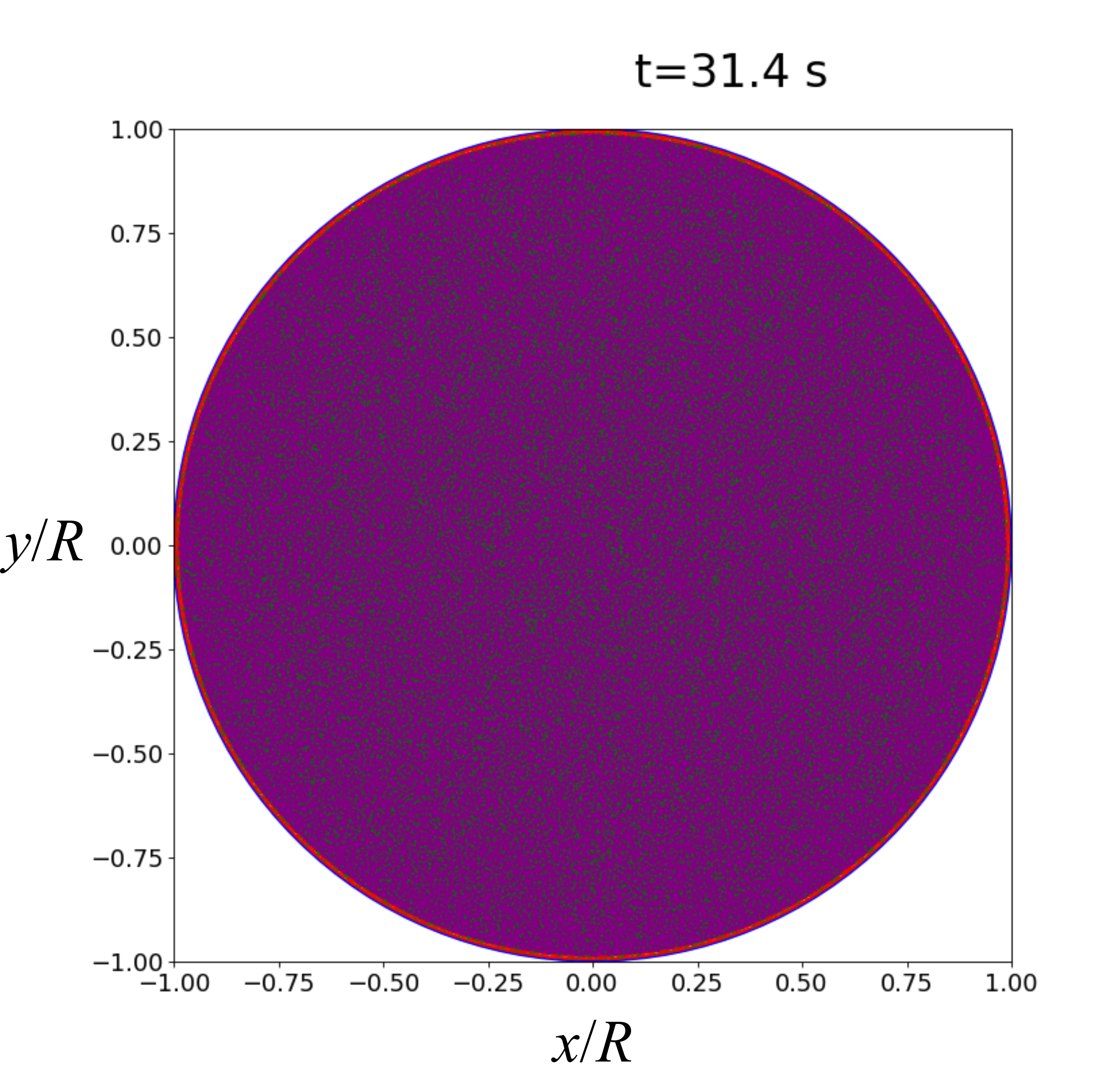}
    \includegraphics[width=0.49\linewidth]{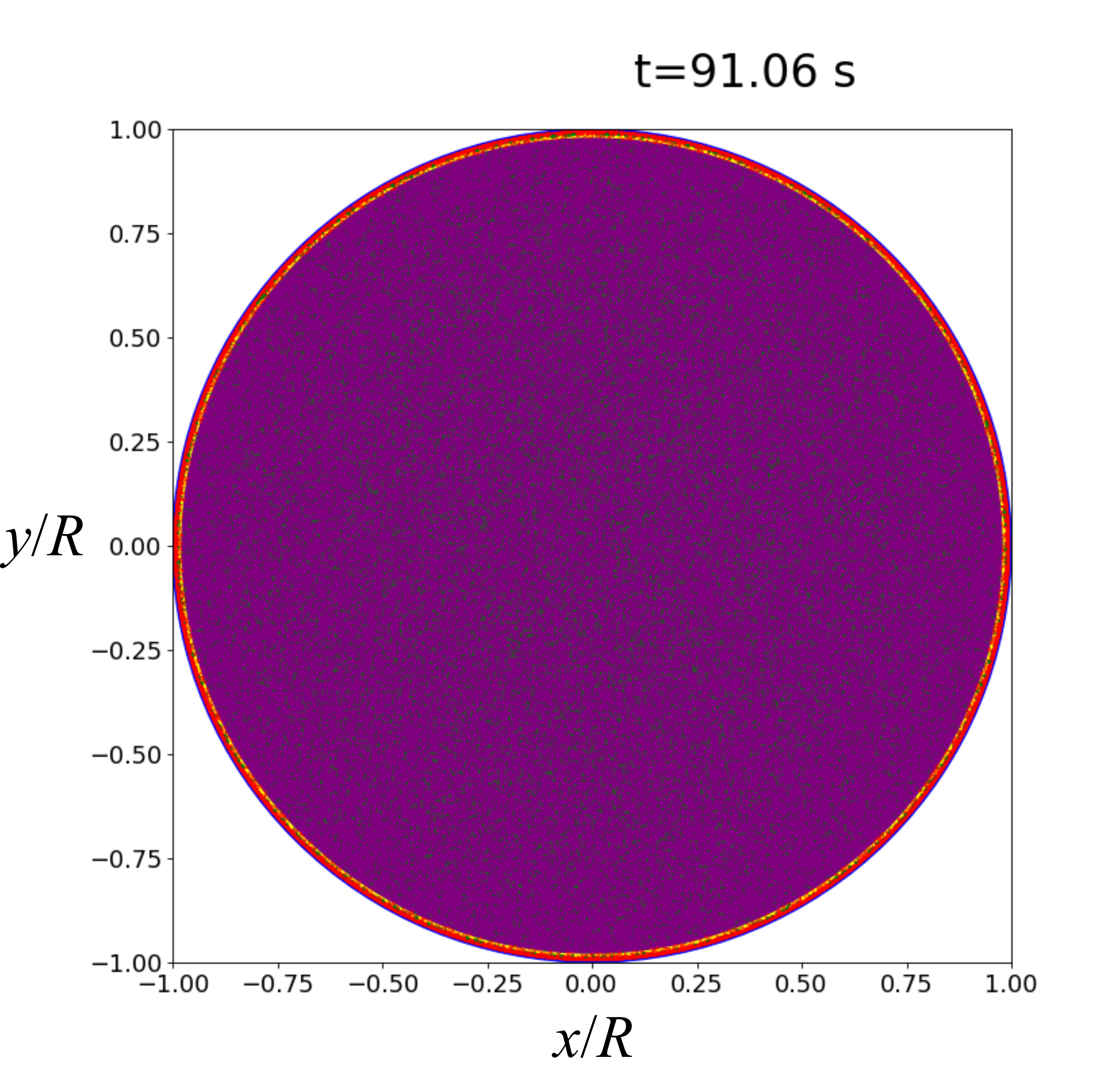}\\
    \includegraphics[width=0.49\linewidth]{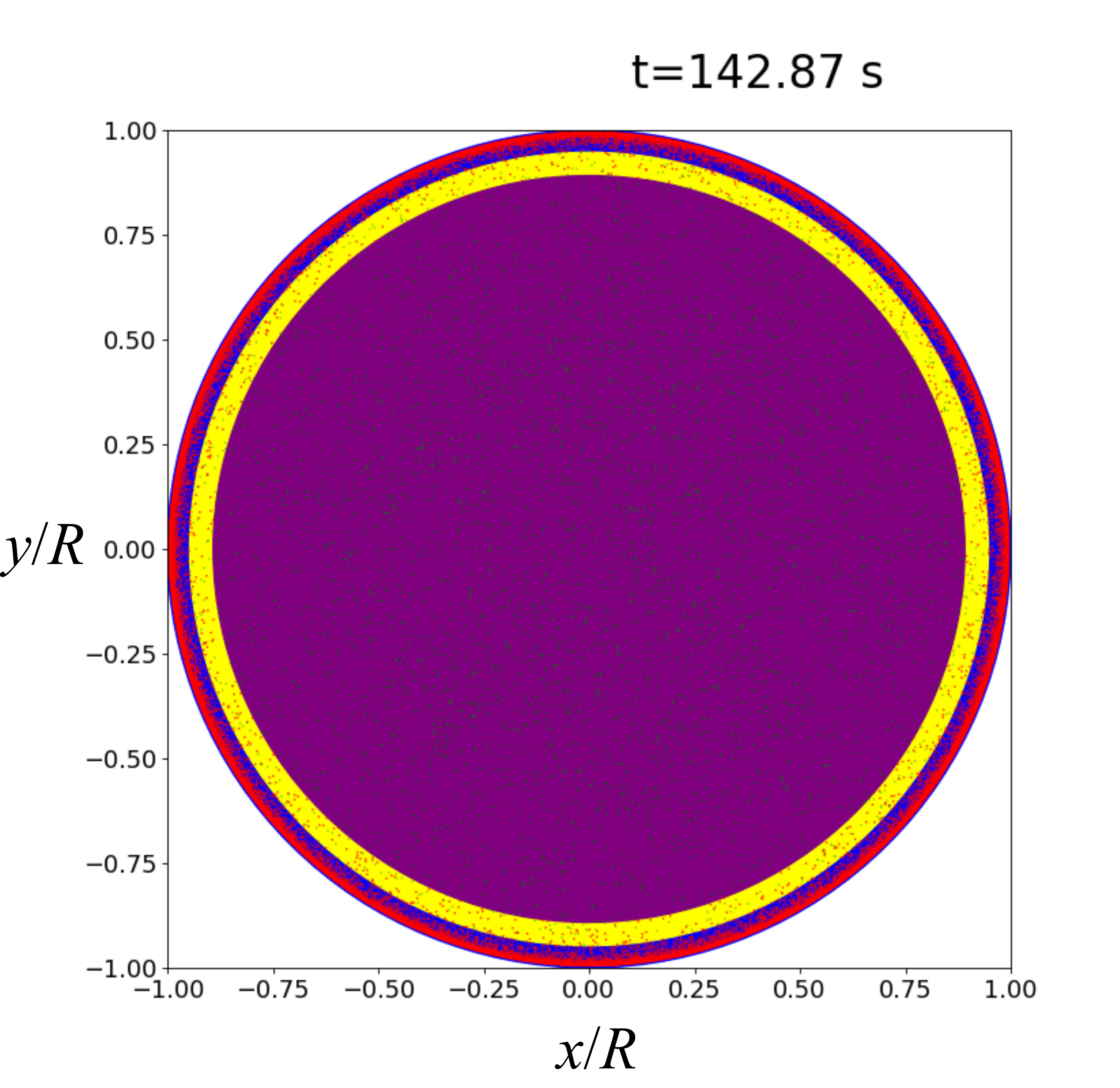}
    \includegraphics[width=0.49\linewidth]{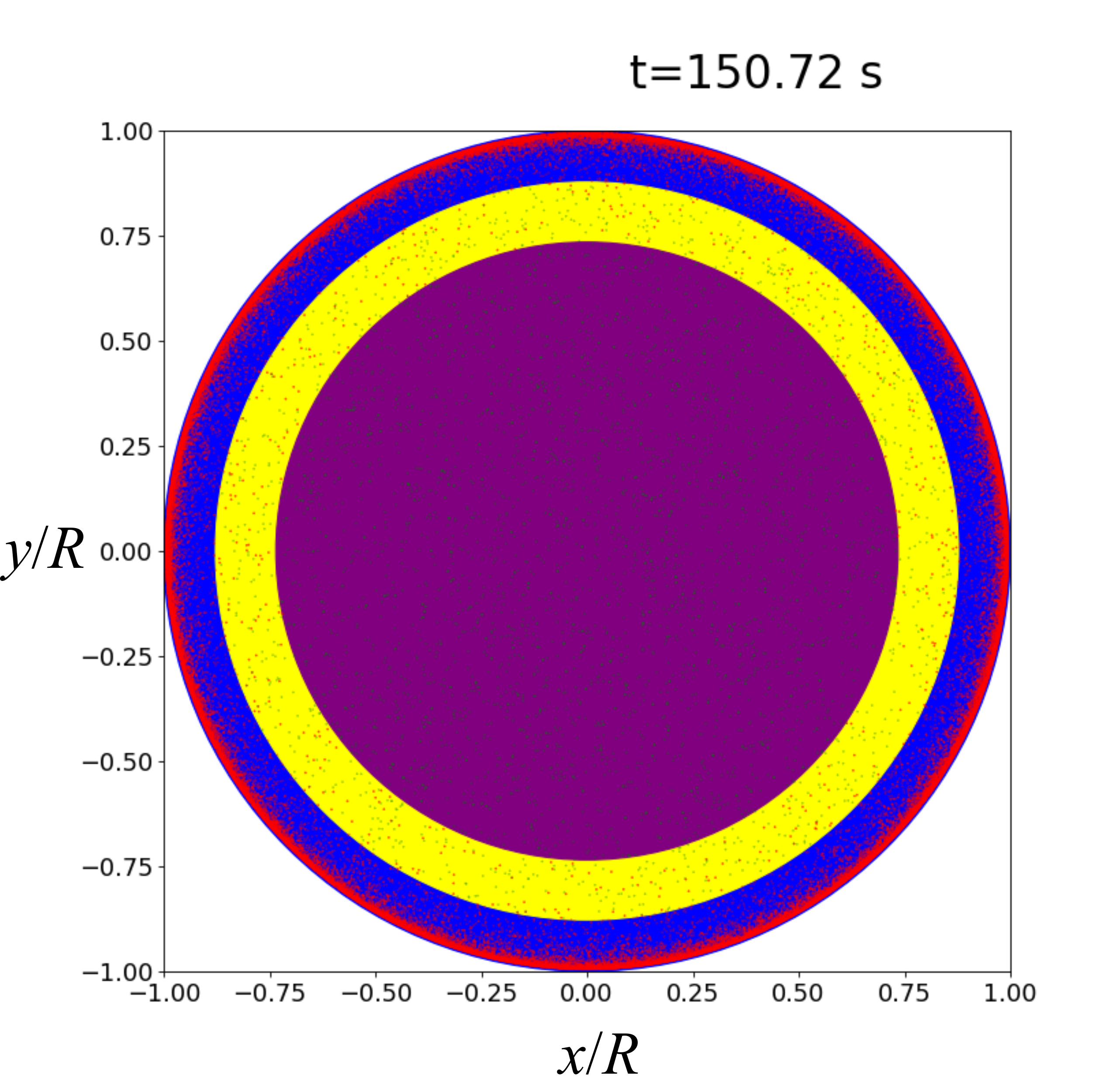}\\
    \includegraphics[width=0.49\linewidth]{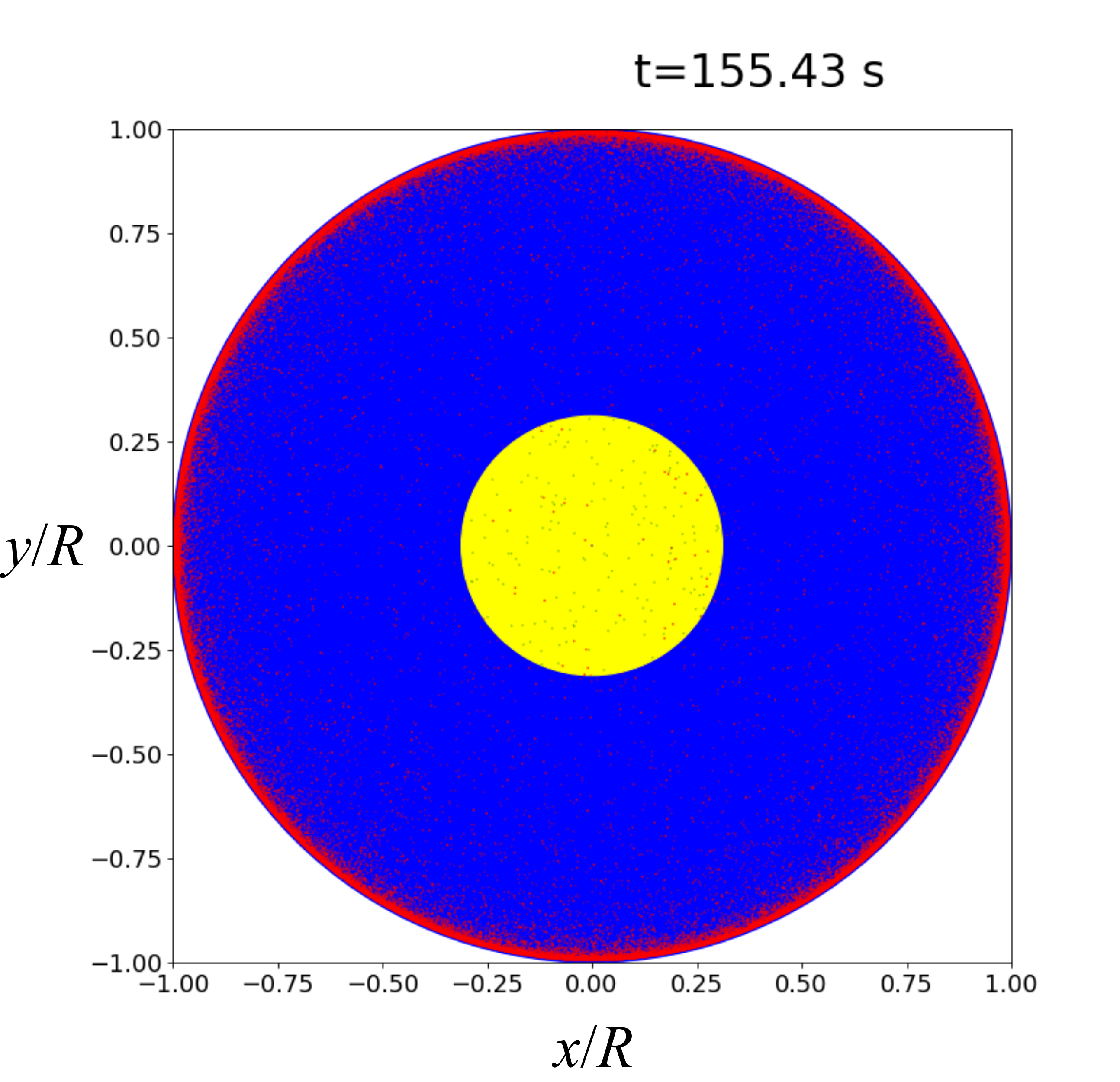}
    \includegraphics[width=0.49\linewidth]{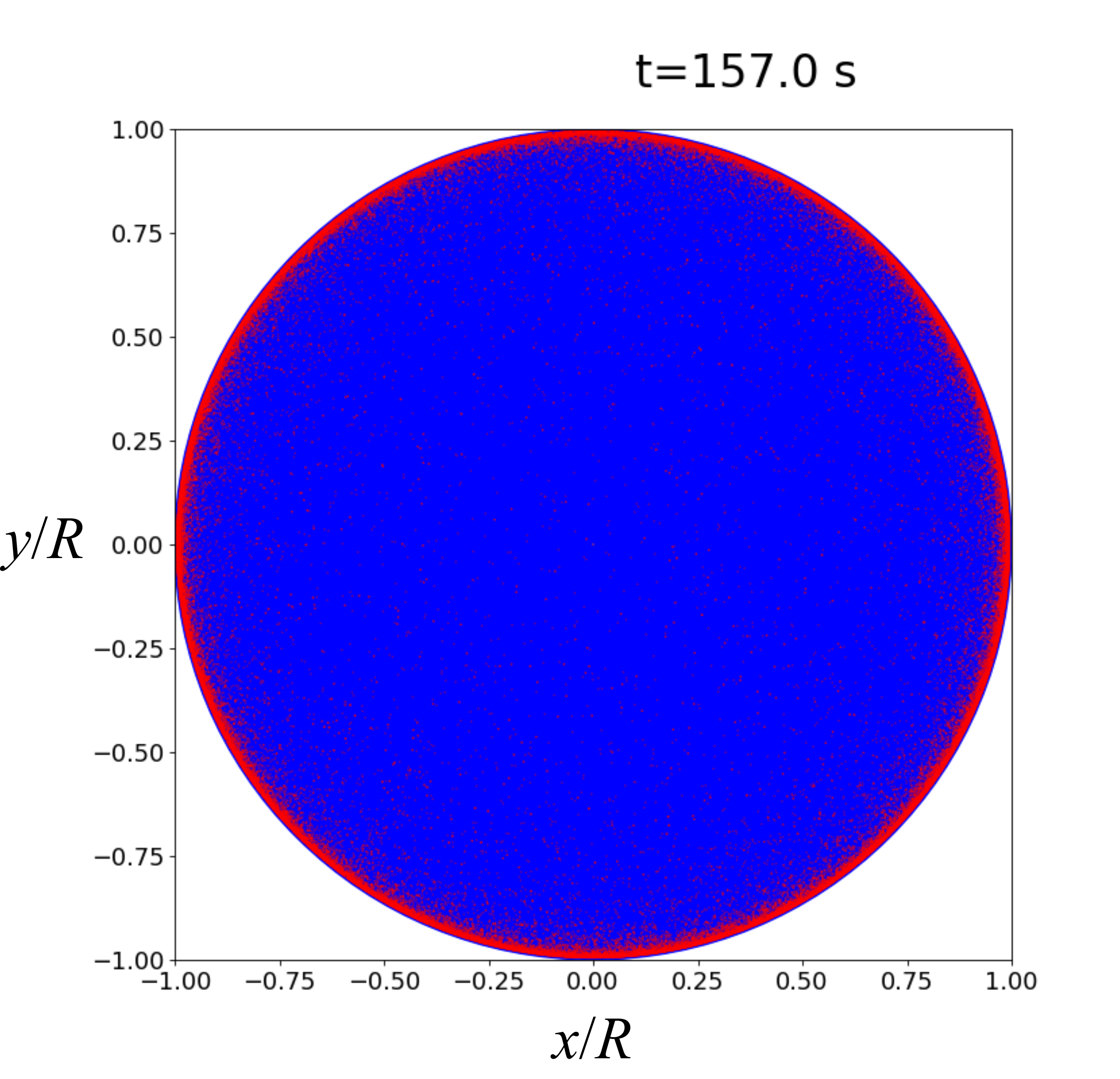}\\
    \caption{
Calculation visualization for several time points at $N_l=$ 7460 and $N_s=$ 59680. Here, the boundary $R_l$ separates the purple and yellow subdomain, and the boundary $R_s$ separates the yellow and blue subdomain. Multimedia view.
    }
    \label{fig:ScreensBi-dispersedProblem}
\end{figure}

The algorithm for the computer program is described in
Algorithm~\ref{alg:particleDynamicsBi-dispersedProblem}. The number
of small, $N_s$, and large particles, $N_l$, is specified. The total
number of particles is $N= N_s + N_l$. An example of visualization
of the calculation is shown in
Fig.~\ref{fig:ScreensBi-dispersedProblem} (multimedia view).
\begin{algorithm}[H]
    \caption{Particle dynamics algorithm}
    \label{alg:particleDynamicsBi-dispersedProblem}
    \begin{algorithmic}[1]
        \State Problem parameters definition: $r_{s,l}$, $R$, $N_{s,l}$, $D_{s,l}$, $\theta_0$, $\delta t$, and $t_\mathrm{max}$.
        \State Generation random coordinates of particles, $(x_i, y_i)$, $i \in [1;N]$.
        \State By default, all particles are marked green.
        \For {$\tau \leftarrow 1, t_\mathrm{max}/\delta t$}
        \State Calculate $R_{s,l}$.
        \For {$i \leftarrow 1, N$}
        \State Change particle status if necessary.
        \EndFor
        \For {$i \leftarrow 1, N$}
        \If {(A current particle is green)}
        \State Calculate new particle coordinates due to diffusion.
        \If {(No collision)}
        \State Move the particle.
        \EndIf
        \State Calculate new particle coordinates due to advection.
        \If {(No collision)}
        \State Move the particle.
        \EndIf
        \EndIf
        \EndFor
        \State Write the particle coordinates and status to a file for the current time step.
        \EndFor
    \end{algorithmic}
\end{algorithm}

Let us mark moving particles in green and stationary particles in red (Fig.~\ref{fig:SketchBi-dispersedProblem}). Green particles are subject to convective and diffusion transfer. By default, all particles are green at the initial moment, $t=0$. The initial distribution of the small and large particles is generated inside circles with radii of $R_s$ or $R_l$, respectively, because a particle cannot be beyond the fixation radius. The coordinates $x$ and $y$ are randomly generated  in a loop for each particle (random sequential adsorption). New values of coordinates  $x$ and $y$ are generated for the current particle if a particle collision occurs (the intersection of circles representing different particles). If a green particle touches the fixation boundary or ends up beyond it ($r^{s,l}\geq R_{s,l}$), the particle becomes fixed and is then marked in red. In other words, the red particles form the deposit. The particle data structure is an array in which information about the coordinates, its radius, status (moving or stationary), and the coordinates of the subdomain containing a particle is stored for each particle. The domain is divided into many square cells (subdomains), the size of which corresponds to the diameter of a large particle. A separate array stores information about all the subdomains and about which particles are in them (particle unique identifiers). This allows us to optimize collision detection, as there is no need to check the potential collision of the current particle with all the others as it can only collide with its neighbors, so it is enough just to scan for potential collisions with these nearby particles. At each time step, an attempt at diffusive and at advective displacement of each green particle is made in turn. All these attempts are accompanied by collision checks. If a collision occurs, the attempt is canceled and is not repeated. In a previous study~\cite{Zolotarev2021}, we compared algorithms involving either single attempts or multiple attempts to displace particles. It was found that, for the used value of $\delta t$, the average number of displacement failures using the different approaches differed by only a couple of percent. Moreover, our estimate showed the distance of diffusion / advective displacement of a particle does not exceed 1\% of its size in any single time step. With such offsets, collisions are unlikely. Thus, with a sufficiently small time step, it is permissible to perform a single attempt at displacement. At predetermined intervals, the program writes information on each particles to a file for the current time. The program was written in C++ language. The calculated data were processed using scripts written in Python.

\section{Results and discussion}
Numerical calculations have been carried out for three particle
concentrations: 1) $ N_l = $ 3730 and $ N_s = $ 29840, 2) $ N_l = $
7460 and $ N_s = $ 59680, 3) $ N_l = $ 14920 and $ N_s = $ 119360.
The ratio of the large and small particles has been chosen so that
the volume fraction of both is the same in the colloidal
solution~\cite{Wong2011}. Each computational experiment was repeated
ten times to establish the statistical error. The simulation results
allow us to observe the dynamics of large and small particles. In
addition, the model predicts the shape and morphology of the deposit
formed after the liquid dries, depending on the specified
parameters.

\begin{figure*}[!htb]
    \includegraphics[width=0.35\linewidth]{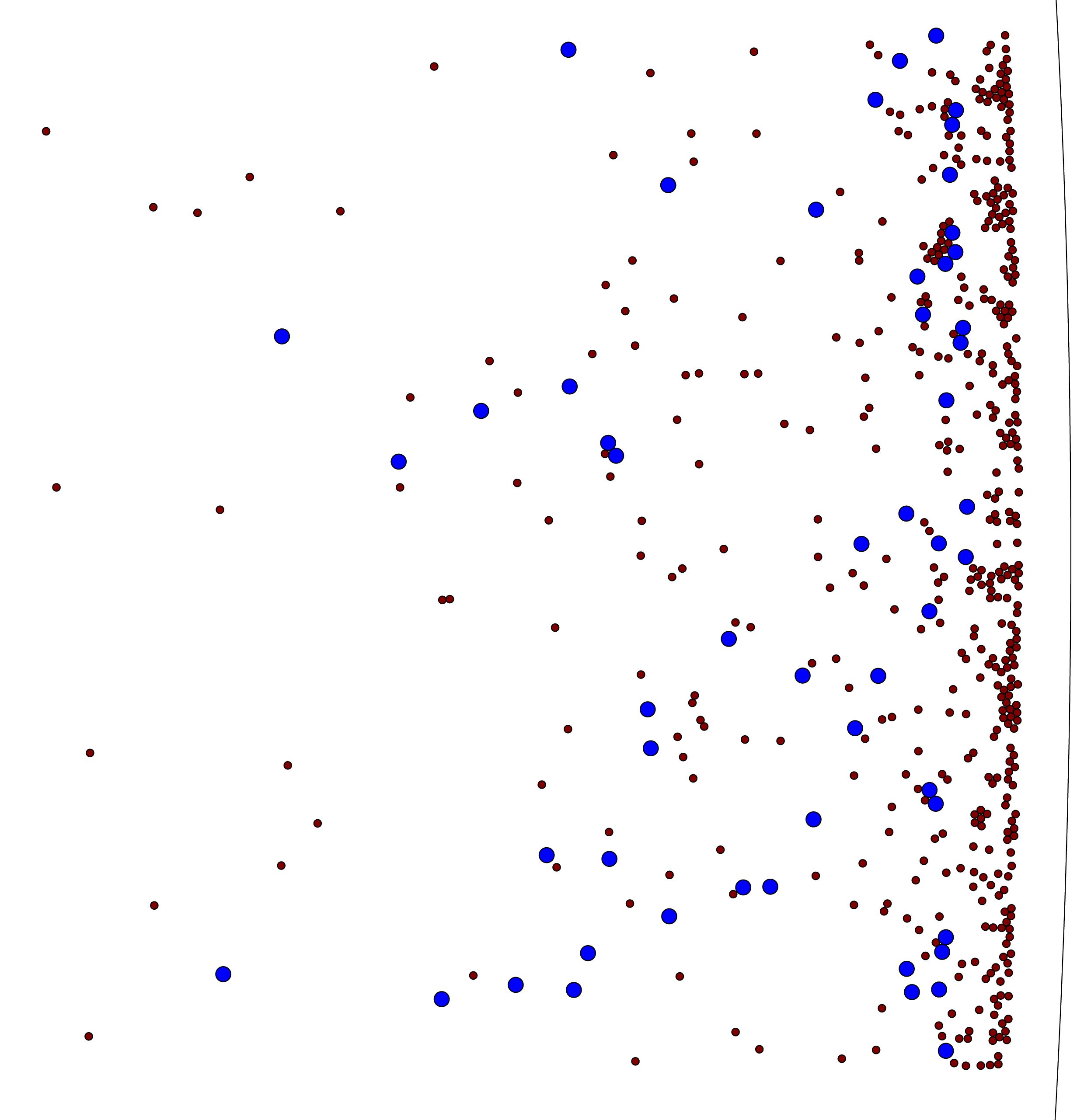} (a) \hspace{15pt}
    \includegraphics[width=0.35\linewidth]{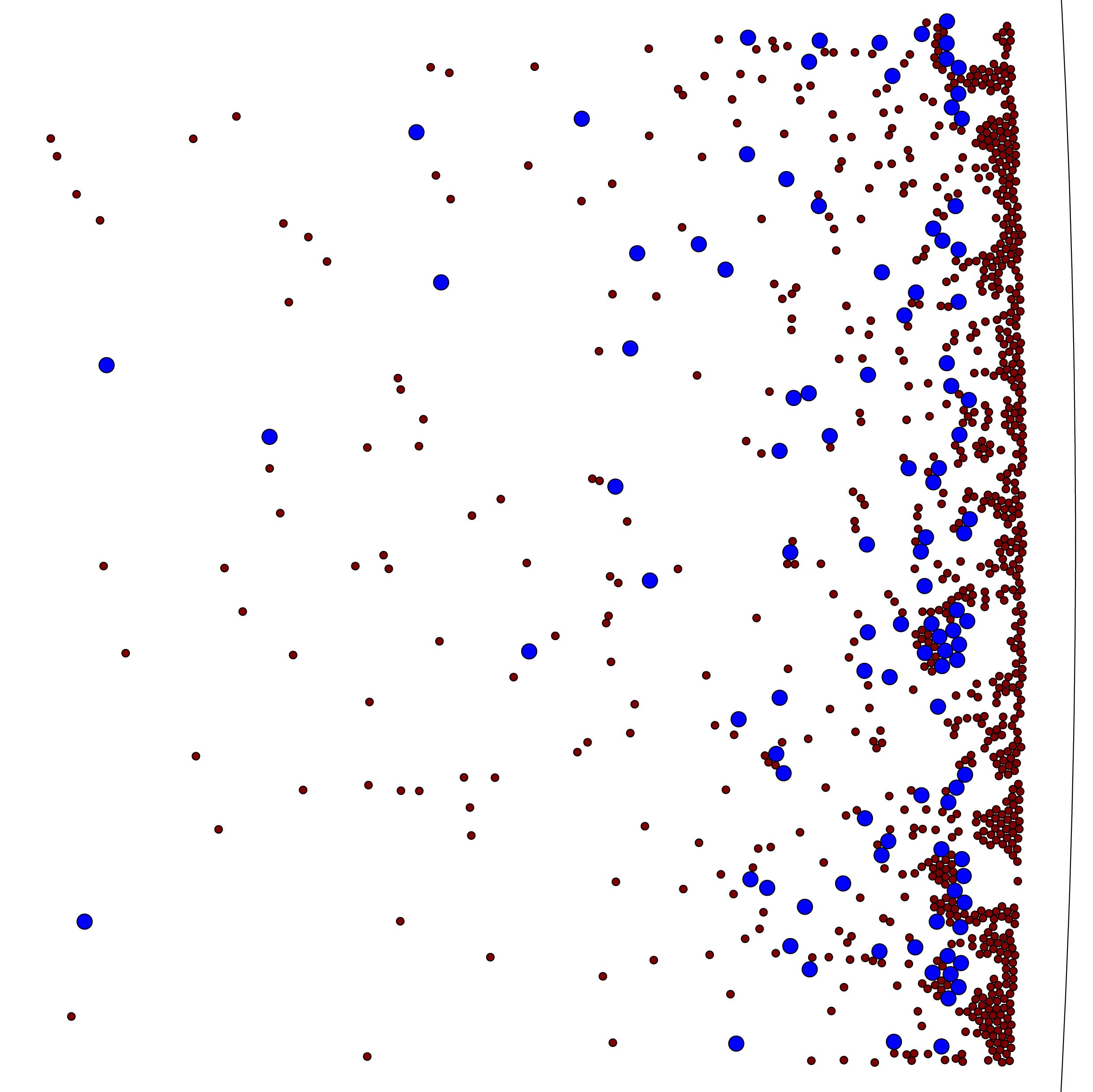} (b)\\
    \includegraphics[width=0.6\linewidth]{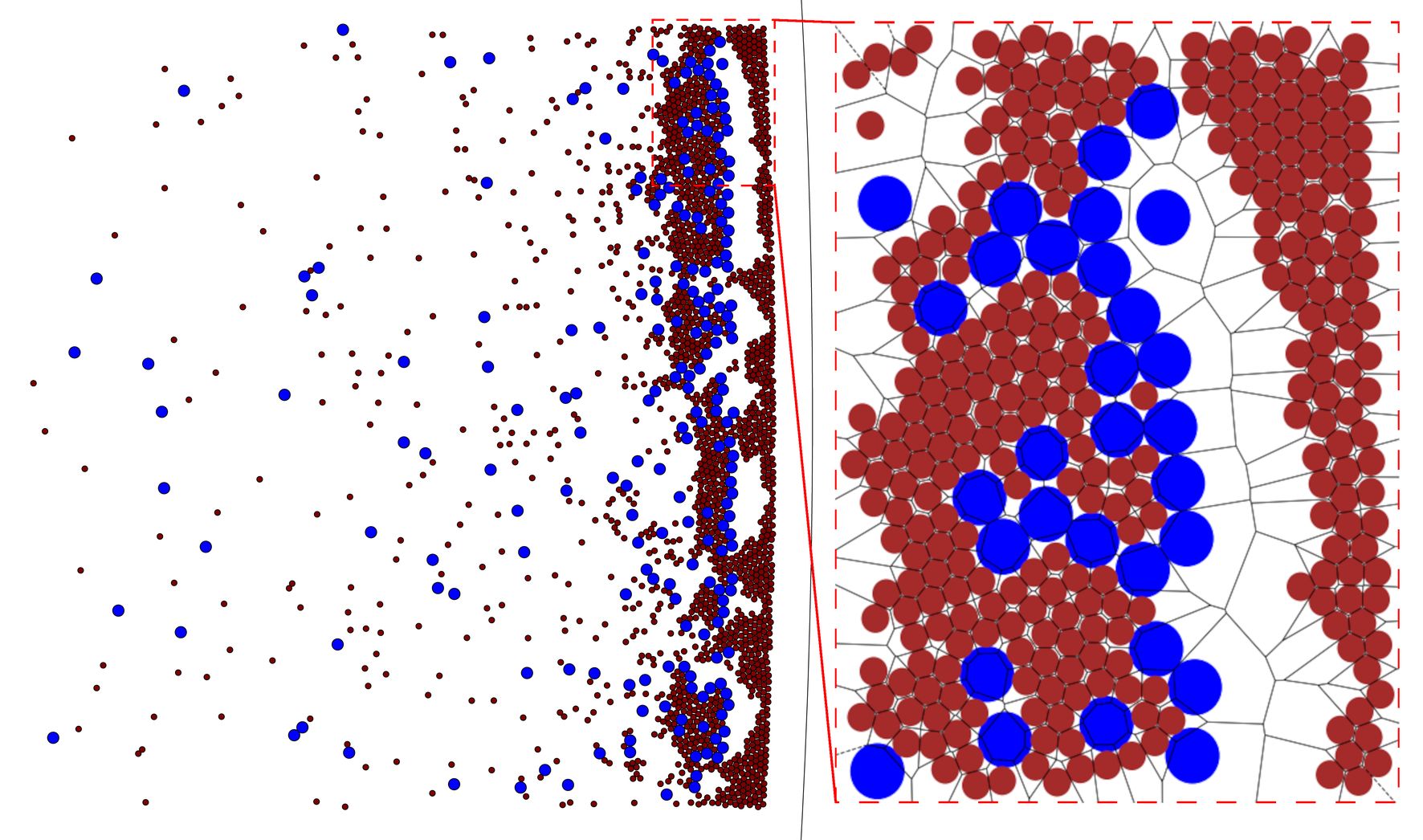} (c)
    \caption{
Final deposit structure near the contact line for different
concentrations: (a) $N_l=$ 3730 and  $N_s=$ 29840, (b) $N_l=$ 7460
and $N_s=$ 59680, (c) $N_l=$ 14920 and $N_s=$ 119360 with Voronoi
diagram inset. }
    \label{fig:sedimentBi-dispersedProblem}
\end{figure*}

\begin{figure}[H]
    \includegraphics[width=0.95\linewidth]{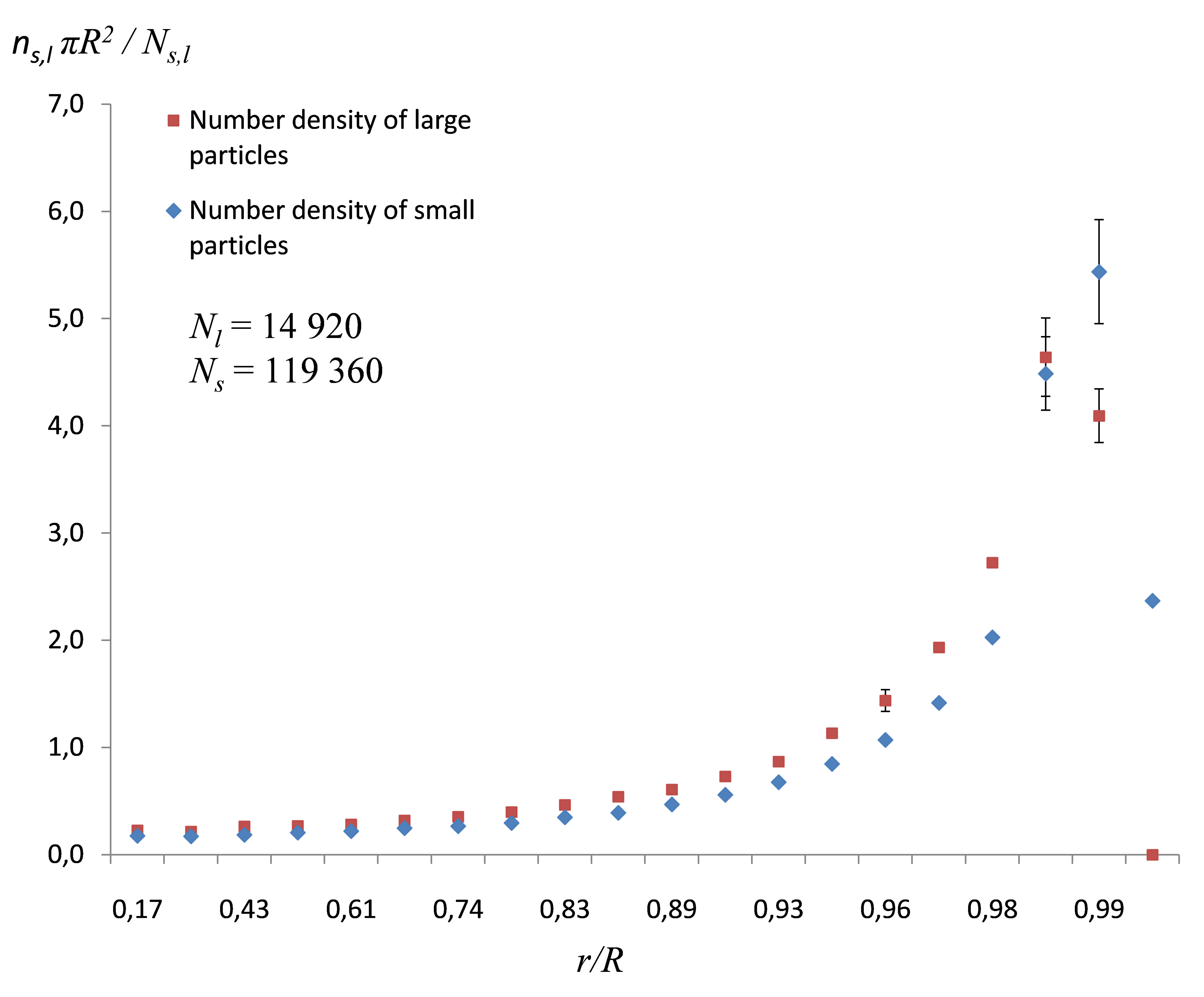}\\
    \caption{
        Scaled number density of the small and large particles for the case with $N_s=$ 119360 and $N_l=$ 14920 at $t=t_\mathrm{max}$ (the error is defined as the standard deviation).
    }
    \label{fig:distributionDensityBi-dispersedProblem}
\end{figure}

Over time, the particles are carried by the flow toward the contact
line (Fig.~\ref{fig:ScreensBi-dispersedProblem}, multimedia view).
Also, they are mixed due to diffusion, including in the area of
annular deposit formation. In the screenshots
(Fig.~\ref{fig:ScreensBi-dispersedProblem}, multimedia view), the
$R_s$ position is between the yellow and blue subareas, and the
$R_l$ position is between the yellow and purple subareas. Moving
green particles are changed to red when they reach the fixation
radii corresponding to their size and stop. At the beginning of the
process, the fixation radii move very slowly. Their movement
gradually accelerates over time, and they rapidly collapse at the
end of the process. The dependence of the fixation radius position
on time was shown in a previous study (Fig.~8 in
Ref.~\cite{Kolegov2019}). At a small distance from the contact line
marked in black (Fig.~\ref{fig:sedimentBi-dispersedProblem}), there
are small particles, while, a little further out, a mixture of large
and small ones can be observed. At relatively high initial
concentrations of a colloidal solution
(Fig.~\ref{fig:sedimentBi-dispersedProblem}b and
\ref{fig:sedimentBi-dispersedProblem}c), particle clusters or
clusters of particles of mixed sizes having extended area are formed
in the deposit near the contact line. In some places, we can observe
a dense hexagonal packing of small particles (see the Voronoi
diagram in Fig.~\ref{fig:sedimentBi-dispersedProblem}c). The methods
for analyzing the morphology of such structures are described in
detail in the reviews~\cite{Lotito2019,Lotito2020}. Narrow and
relatively free spaces are visible between mixed particle clusters
and clusters of small particles
(Fig.~\ref{fig:sedimentBi-dispersedProblem}b and
\ref{fig:sedimentBi-dispersedProblem}c). In the case of a low
concentration of the colloidal solution, particle clusters are
either small or non-existent
(Fig.~\ref{fig:sedimentBi-dispersedProblem}a). The number of
particles in the deposit per unit area (number density),
$n_{s,l}=\bar N_{s,l}/S_\mathrm{ring}$, has been calculated, where
$S_\mathrm{ring}$ is a ring square and $\bar N_{s,l}$ is the number
of  local particles. The region is divided into concentric rings,
the width of which increases linearly towards the center of the
drop, since there are fewer particles there by the end of the
process. Each ring contains a local number of particles $\bar
N_{s,l}$. The dependence of the number density on the spatial
coordinate $r$ shows that a larger number of particles are located
on the periphery of the dried drop
(Fig.~\ref{fig:distributionDensityBi-dispersedProblem}).

The $n_l$ value in the entire region slightly exceeds $n_s$ except for the region near the contact line. The situation reverses here ($n_s > n_l$) since large particles cannot approach closer to the contact line than the small ones. Fig.~\ref{fig:distributionDensityBi-dispersedProblem} shows a graph for only one case, in which $N_l=$ 14920 and $N_s=$ 119360. However, for the other two cases, the number density does not differ qualitatively (see the Supplement).

In the experiment, the distance between the outermost ring of nanoparticles and the innermost ring of microparticles, $\Delta L$, increased if the particle volume fraction was lower than a critical value~\cite{Wong2011}. The number of nanoparticles is orders of magnitude greater than microparticles at the same volume fraction. Therefore, the surface tension force does not cause the nanoparticle sediment to shift towards the center of the droplet. However, the microparticle layer shifts slightly toward the central region until an equilibrium is reached between the surface tension force and the particle-substrate adhesion force~\cite{Jung2009,Jung2010}. Adhesion prevails over surface tension at a concentration above the critical value so the distance $\Delta L$ can be determined from geometric considerations~\cite{Wong2011}. Although this effect should be taken into account in future detailed modeling, it does not need to be  considered in the proposed simple model. Most likely, a dissipative particle dynamics approach is better suited to this~\cite{LebedevStepanov2013}. In the Langevin equation describing the motion of particles, it would be possible to take into account additional forces, including adhesion~\cite{Jung2009,Jung2010}. In the model used here, the maximum distance between the inner and outer rings can be calculated using a theoretical formula $\Delta L \approx (d_l-d_s) / \theta_0 \approx R_s(0)-R_l(0)$ at $\theta_0 \to 0$. The distance between the outer ring and the three-phase boundary is $\Delta l \approx d_s / \theta_0 \approx R-R_s(0)$. After substituting the values of the parameters, we get  $\Delta L \approx \Delta l \approx$ 6~$\mu$m. Here, we have considered the case with the ratio $\gamma = r_s / r_l = 0.5$. The parameter $\gamma$ affects the distance between the rings. In the future, it will be necessary to develop a 3D model in order to consider the case of $\gamma \ll 1$.

We have determined and analyzed some dimensionless parameters
associated with annular deposits based on our calculated data
(Table~\ref{tab:NondimensionalParameters}). The main problem is to
determine the inner radii of annular precipitation. Let us denote
the inner radius of the inner ring as $\check R$ and the inner
radius of the outer ring as $\hat R$. The area of the outer annular
deposit is $\hat S=\pi (R_s^2(0)-\hat R^2)$. We assume that the
particle packing density corresponds to a random packing,
$p_r\approx 0.64$. On the other hand, we have the area $\hat
S\approx \hat N_s \pi r_s^2/ p_r$, where $\hat N_s$ is the number of
small particles in the outer annular deposit. The value of $\hat
N_s$ can be calculated using the condition of $r>R_l(0)$. Thus, we
have obtained the expression $p_r (R_s^2(0)-\hat R^2)\approx \hat
N_s r_s^2$, from which the unknown parameter $\hat R$ can be found,
$$\hat R\approx \sqrt{R_s^2(0)-\frac{N_s r_s^2}{p_r}}.$$ The width
of the outer annular deposit is $\hat w\approx R_s(0)-\hat R$. In
the dimensionless form, we have obtained $\tilde w_\mathrm{out}=
\hat w/ \Delta L$, where $\Delta L\approx R_s(0)-R_l(0)$ and $0\leq
\tilde w_\mathrm{out}\leq 1$. Assume that the number of particles in
the inner annular deposit is $\check N\approx (N_s-\hat N_s)+N_l$.
This amount is overestimated since some of the particles are
deposited in the central region. The area of the inner annular
deposit is expressed as $\check S= \pi (R_l^2(0)-\check R^2)$, but,
on the other hand, we have $\check S\approx ( (N_s-\hat N_s) r_s^2 +
N_l r_l^2 )\pi/ p_r$. Taking into account these formulas, we have
obtained $\check R$,
$$\check R\approx \sqrt{R_l^2(0)-\frac{(N_s-\hat N_s) r_s^2 + N_l r_l^2}{p_r}}.$$ Hence, the width of the inner ring is $\check w\approx R_l(0)-\check R$. In dimensionless form, we have $\tilde w_\mathrm{in}= \check w/ \Delta L$, where $\tilde w_\mathrm{in} \geq
 0$.
\begin{table}[h]
    \caption{Nondimensional parameters}
    \centering
    \begin{tabular}{|p{0.28\linewidth}|p{0.21\linewidth}|p{0.21\linewidth}|p{0.21\linewidth}|}
        \hline
        Parameters~/ number of particles & $N_l=3730$, $N_s= 29840$ &  $N_l=7460$, $N_s=59680$ & $N_l=14920$, $N_s=119360$ \\
        \hline
        $\hat p$ & 0.295 & 0.486 & 0.608 \\
        $\check p$ & 0.156 & 0.277 & 0.414 \\
        $\hat N_s/N_s$ & 0.229 & 0.35 & 0.313 \\
        $(\check N_s + \check N_l)/N$ & 0.229 & 0.35 & 0.313 \\
        $(\check N_s + \check N_l)/N$ & 0.138 & 0.253 & 0.469 \\
        $\check A_n$ & 1.029 & 1.936 & 3.219 \\
        $\check a_n$ & 0.549 & 1.488 & 2.816 \\
        $\hat a_n$ & 1.16 & 2.418 & 3.282 \\
        $\tilde w_\mathrm{in}$ & 0.714 & 1.47 & 3.325 \\
        $\tilde w_\mathrm{out}$ & 0.342 & 0.644 & 0.922 \\
        \hline
    \end{tabular}
    \label{tab:NondimensionalParameters}
\end{table}

The packing densities of particles are calculated in the inner,
$\check p$, and in the outer ring, $\hat p$, deposits as the ratio
of the area covered by particles in a ring to the area of the ring
itself. According to Table~\ref{tab:NondimensionalParameters},
values of $\check p$ and of $\hat p$ increase with an increase in
the initial concentration of particles in the colloidal solution.
When calculating $\hat p$, the pieces of the large particles
slightly protruding through the boundary of $R_l(0)$ are also taken
into account, but their coordinate is $r\leq R_l(0)$. At first, the
ratio of the number of the small particles in the outer ring to the
total number of the small particles, $\hat N_s/N_s$, also increases
with the increase in concentration but then it decreases. This may
be due to a high concentration, at which the large particles, near
their fixation radius, begin to prevent the small particles from
advancing towards the periphery of the droplet. This effect leads to
the increase in the ratio of the particle number in the inner ring
to the total number of particles, $(\check N_s + \check N_l)/N$,
with the increase in initial concentration. The average number of
neighbors of each particle in the inner and outer rings also
increases with increasing concentration. In our calculations, two
particles were considered adjacent if the condition $d\leq
\varepsilon(r_1+r_2)$ was true. Here, $r_1$ and $r_2$ are the radii
of two relatively close particles located in the same subdomain
(cell) or in adjacent cells, $d$ is the distance between two
particles, and the parameter $\varepsilon= 1.01$ is used to handle
the round-off errors with floating-point. It should be noted that
for all the concentrations considered, the average number of
neighbors for large particles $\check A_n$ is slightly higher than
the value for small particles $\check a_n$ in the inner ring. We
have found that the ratio is of $\check A_n/ \check a_n\approx 1.87$
for a low concentration (case 1), of $\check A_n/ \check a_n\approx
1.3$ for a moderate concentration  (case 2), and of $\check A_n/
\check a_n\approx 1.14$ for a high concentration  (case 3). Also, we
should note that, for all the concentrations considered, the average
number of neighbors of small particles in the outer ring $\hat a_n$
is slightly higher than the corresponding values for small particles
in the inner ring, $\hat a_n > \check a_n$. As the concentration
increases, we observe the ratios $\hat a_n / \check a_n\approx
2.11$, 1.63, and 1.16. Thus, we conclude that, as the concentration
increases, the ratios $\check A_n/ \check a_n\to 1$ and $\hat a_n /
\check a_n\to 1$ are expected. This is due to an increase in the
packing densities $\check p$ and $\hat p$ with an increase in the
concentration of the colloidal solution. In addition, the higher the
concentration, the greater the width of both rings formed are
observed. For the largest particle number considered here, the width
of the outer ring $\tilde w_\mathrm{out}$ is close to the maximum
value corresponding to the distance between the two fixation radii,
$\Delta L$. The width of the inner ring can be less than $\Delta L$
or several times greater than $\Delta L$, depending on the number of
particles in the system. With increasing concentration, the ratio of
the width of the inner ring to the outer one increases, $\tilde
w_\mathrm{in}/ \tilde w_\mathrm{out}\approx 2.1$, 2.3, and 3.6.
Also, this is probably since large particles can interfere with
small ones in their further movement to the edge of the area. The
more large particles, the more such obstacles occur.

\section{Conclusion}

An important and actively discussed problem is that of studying the structures of colloidal particles that form on a substrate after sessile droplet evaporation. One such example is the effect of deposition near a contact line during evaporation. This results in the so-called coffee-ring effect. While a droplet is drying on the substrate, capillary flows carry the colloid particles toward the three-phase boundary. In this case, if the contact line is pinned throughout the entire process, the formation of an annular deposition is observed. Tak-Sing Wong et al.~\cite{Wong2011} showed in their experiment that it is possible to use the coffee ring effect for separating suspended particles by their size (Fig.~\ref{fig:WongExperiment}). This is useful for a variety of applications. For example, it has direct implications for developing low-cost technologies for disease diagnostics in resource-poor environments. Thus, understanding the mechanism of particle separation near the three-phase boundary plays an important role in the further development of specific applications.

The model~\cite{Devlin2015} did not provide an explanation for the effect of particle separation according to size. The results of those calculations predicted the accumulation of large particles near the contact line, which contradicts experimental observations~\cite{Devlin2015}. Here, we have proposed a new method for modeling this phenomenon. Our simple model, taking into account advective transport, particle diffusion, and the specific geometry of the region near the three-phase boundary, has allowed us to obtain numerical results, that are in qualitative agreement with the experimental ones~\cite{Wong2011}. The simulation results have shown that the particles do not reach the contact line, but accumulate at a small distance from it. The reason for this is the surface tension acting on the particles in areas where the thickness of the liquid layer is comparable to their size. The same mechanism affects the separation of small and large particles. Large particles deposit at a short distance from the clusters of small particles, along with some small particles that are prevented from moving even closer to the contact line because of obstruction by the large ones. An increase in the number of large particles in the system can lead to a decrease in the ratio of small particles, $\hat N_s/ N_s$, reaching the outer annular deposit. Obstacles arising from large particles at a relatively high concentration of colloidal solution also lead to an increase in the ratio of the width of the inner ring to the width of the outer one, $\tilde w_\mathrm{in}/ \tilde w_\mathrm{out}$. This model is phenomenological. To describe the process quantitatively, more complicated and more accurate models need to be developed.

\section{Supplementary Material}
See the supplementary material for showing scaled number density of the small and large particles for the case with $N_s=$ 29840, $N_l=$ 3730 and with $N_s=$ 59680, $N_l=$ 7460 at $t=t_\mathrm{max}$.

\begin{acknowledgments}
The project (GK21-001093) is being implemented by the winner of the Master's program faculty grant competition 2020/2021 of the Vladimir Potanin fellowship program. The authors thank prof. Yuri Tarasevich for his useful comments.
\end{acknowledgments}

\section*{Data Availability Statement}

The data that support the findings of this study are available
from the corresponding author upon reasonable request.

\nocite{*}
\bibliography{biblio}

\begin{thebibliography}{67}%
\makeatletter
\providecommand \@ifxundefined [1]{%
 \@ifx{#1\undefined}
}%
\providecommand \@ifnum [1]{%
 \ifnum #1\expandafter \@firstoftwo
 \else \expandafter \@secondoftwo
 \fi
}%
\providecommand \@ifx [1]{%
 \ifx #1\expandafter \@firstoftwo
 \else \expandafter \@secondoftwo
 \fi
}%
\providecommand \natexlab [1]{#1}%
\providecommand \enquote  [1]{``#1''}%
\providecommand \bibnamefont  [1]{#1}%
\providecommand \bibfnamefont [1]{#1}%
\providecommand \citenamefont [1]{#1}%
\providecommand \href@noop [0]{\@secondoftwo}%
\providecommand \href [0]{\begingroup \@sanitize@url \@href}%
\providecommand \@href[1]{\@@startlink{#1}\@@href}%
\providecommand \@@href[1]{\endgroup#1\@@endlink}%
\providecommand \@sanitize@url [0]{\catcode `\\12\catcode `\$12\catcode
  `\&12\catcode `\#12\catcode `\^12\catcode `\_12\catcode `\%12\relax}%
\providecommand \@@startlink[1]{}%
\providecommand \@@endlink[0]{}%
\providecommand \url  [0]{\begingroup\@sanitize@url \@url }%
\providecommand \@url [1]{\endgroup\@href {#1}{\urlprefix }}%
\providecommand \urlprefix  [0]{URL }%
\providecommand \Eprint [0]{\href }%
\providecommand \doibase [0]{https://doi.org/}%
\providecommand \selectlanguage [0]{\@gobble}%
\providecommand \bibinfo  [0]{\@secondoftwo}%
\providecommand \bibfield  [0]{\@secondoftwo}%
\providecommand \translation [1]{[#1]}%
\providecommand \BibitemOpen [0]{}%
\providecommand \bibitemStop [0]{}%
\providecommand \bibitemNoStop [0]{.\EOS\space}%
\providecommand \EOS [0]{\spacefactor3000\relax}%
\providecommand \BibitemShut  [1]{\csname bibitem#1\endcsname}%
\let\auto@bib@innerbib\@empty
\bibitem [{\citenamefont {Harris}, \citenamefont {Conrad},\ and\ \citenamefont
  {Lewis}(2009)}]{Harris2009}%
  \BibitemOpen
  \bibfield  {author} {\bibinfo {author} {\bibfnamefont {D.~J.}\ \bibnamefont
  {Harris}}, \bibinfo {author} {\bibfnamefont {J.~C.}\ \bibnamefont {Conrad}},\
  and\ \bibinfo {author} {\bibfnamefont {J.~A.}\ \bibnamefont {Lewis}},\
  }\bibfield  {title} {\enquote {\bibinfo {title} {Evaporative lithographic
  patterning of binary colloidal films},}\ }\href
  {https://doi.org/10.1098/rsta.2009.0157} {\bibfield  {journal} {\bibinfo
  {journal} {Philosophical Transactions of the Royal Society A: Mathematical,
  Physical and Engineering Sciences}\ }\textbf {\bibinfo {volume} {367}},\
  \bibinfo {pages} {5157--5165} (\bibinfo {year} {2009})}\BibitemShut {NoStop}%
\bibitem [{\citenamefont {Utgenannt}\ \emph {et~al.}(2013)\citenamefont
  {Utgenannt}, \citenamefont {Keddie}, \citenamefont {Muskens},\ and\
  \citenamefont {Kanaras}}]{Utgenannt2013}%
  \BibitemOpen
  \bibfield  {author} {\bibinfo {author} {\bibfnamefont {A.}~\bibnamefont
  {Utgenannt}}, \bibinfo {author} {\bibfnamefont {J.~L.}\ \bibnamefont
  {Keddie}}, \bibinfo {author} {\bibfnamefont {O.~L.}\ \bibnamefont
  {Muskens}},\ and\ \bibinfo {author} {\bibfnamefont {A.~G.}\ \bibnamefont
  {Kanaras}},\ }\bibfield  {title} {\enquote {\bibinfo {title} {Directed
  organization of gold nanoparticles in polymer coatings through
  infrared-assisted evaporative lithography},}\ }\href
  {https://doi.org/10.1039/c2cc37844b} {\bibfield  {journal} {\bibinfo
  {journal} {Chemical Communications}\ }\textbf {\bibinfo {volume} {49}},\
  \bibinfo {pages} {4253--4255} (\bibinfo {year} {2013})}\BibitemShut {NoStop}%
\bibitem [{\citenamefont {Kolegov}\ and\ \citenamefont
  {Barash}(2020)}]{Kolegov2020}%
  \BibitemOpen
  \bibfield  {author} {\bibinfo {author} {\bibfnamefont {K.~S.}\ \bibnamefont
  {Kolegov}}\ and\ \bibinfo {author} {\bibfnamefont {L.~Y.}\ \bibnamefont
  {Barash}},\ }\bibfield  {title} {\enquote {\bibinfo {title} {Applying
  droplets and films in evaporative lithography},}\ }\href
  {https://doi.org/10.1016/j.cis.2020.102271} {\bibfield  {journal} {\bibinfo
  {journal} {Advances in Colloid and Interface Science}\ }\textbf {\bibinfo
  {volume} {285}},\ \bibinfo {pages} {102271} (\bibinfo {year}
  {2020})}\BibitemShut {NoStop}%
\bibitem [{\citenamefont {Al-Muzaiqer}\ \emph {et~al.}(2021)\citenamefont
  {Al-Muzaiqer}, \citenamefont {Kolegov}, \citenamefont {Ivanova},\ and\
  \citenamefont {Fliagin}}]{AlMuzaiqer2021}%
  \BibitemOpen
  \bibfield  {author} {\bibinfo {author} {\bibfnamefont {M.~A.}\ \bibnamefont
  {Al-Muzaiqer}}, \bibinfo {author} {\bibfnamefont {K.~S.}\ \bibnamefont
  {Kolegov}}, \bibinfo {author} {\bibfnamefont {N.~A.}\ \bibnamefont
  {Ivanova}},\ and\ \bibinfo {author} {\bibfnamefont {V.~M.}\ \bibnamefont
  {Fliagin}},\ }\bibfield  {title} {\enquote {\bibinfo {title} {Nonuniform
  heating of a substrate in evaporative lithography},}\ }\href
  {https://doi.org/10.1063/5.0061713} {\bibfield  {journal} {\bibinfo
  {journal} {Physics of Fluids}\ }\textbf {\bibinfo {volume} {33}},\ \bibinfo
  {pages} {092101} (\bibinfo {year} {2021})}\BibitemShut {NoStop}%
\bibitem [{\citenamefont {Trantum}, \citenamefont {Wright},\ and\ \citenamefont
  {Haselton}(2011)}]{Trantum2011}%
  \BibitemOpen
  \bibfield  {author} {\bibinfo {author} {\bibfnamefont {J.~R.}\ \bibnamefont
  {Trantum}}, \bibinfo {author} {\bibfnamefont {D.~W.}\ \bibnamefont
  {Wright}},\ and\ \bibinfo {author} {\bibfnamefont {F.~R.}\ \bibnamefont
  {Haselton}},\ }\bibfield  {title} {\enquote {\bibinfo {title}
  {Biomarker-mediated disruption of coffee-ring formation as a low resource
  diagnostic indicator},}\ }\href {https://doi.org/10.1021/la203903a}
  {\bibfield  {journal} {\bibinfo  {journal} {Langmuir}\ }\textbf {\bibinfo
  {volume} {28}},\ \bibinfo {pages} {2187--2193} (\bibinfo {year}
  {2011})}\BibitemShut {NoStop}%
\bibitem [{\citenamefont {Rathaur}\ \emph {et~al.}(2020)\citenamefont
  {Rathaur}, \citenamefont {Kumar}, \citenamefont {Panigrahi},\ and\
  \citenamefont {Panda}}]{Rathaur2020}%
  \BibitemOpen
  \bibfield  {author} {\bibinfo {author} {\bibfnamefont {V.~S.}\ \bibnamefont
  {Rathaur}}, \bibinfo {author} {\bibfnamefont {S.}~\bibnamefont {Kumar}},
  \bibinfo {author} {\bibfnamefont {P.~K.}\ \bibnamefont {Panigrahi}},\ and\
  \bibinfo {author} {\bibfnamefont {S.}~\bibnamefont {Panda}},\ }\bibfield
  {title} {\enquote {\bibinfo {title} {Investigating the effect of
  antibody{\textendash}antigen reactions on the internal convection in a
  sessile droplet via microparticle image velocimetry and {DLVO} analysis},}\
  }\href {https://doi.org/10.1021/acs.langmuir.0c01162} {\bibfield  {journal}
  {\bibinfo  {journal} {Langmuir}\ }\textbf {\bibinfo {volume} {36}},\ \bibinfo
  {pages} {8826--8838} (\bibinfo {year} {2020})}\BibitemShut {NoStop}%
\bibitem [{\citenamefont {Liu}\ \emph {et~al.}(2019)\citenamefont {Liu},
  \citenamefont {Midya}, \citenamefont {Kappl}, \citenamefont {Butt},\ and\
  \citenamefont {Nikoubashman}}]{Liu2019}%
  \BibitemOpen
  \bibfield  {author} {\bibinfo {author} {\bibfnamefont {W.}~\bibnamefont
  {Liu}}, \bibinfo {author} {\bibfnamefont {J.}~\bibnamefont {Midya}}, \bibinfo
  {author} {\bibfnamefont {M.}~\bibnamefont {Kappl}}, \bibinfo {author}
  {\bibfnamefont {H.-J.}\ \bibnamefont {Butt}},\ and\ \bibinfo {author}
  {\bibfnamefont {A.}~\bibnamefont {Nikoubashman}},\ }\bibfield  {title}
  {\enquote {\bibinfo {title} {Segregation in drying binary colloidal
  droplets},}\ }\href {https://doi.org/10.1021/acsnano.9b00459} {\bibfield
  {journal} {\bibinfo  {journal} {{ACS} Nano}\ }\textbf {\bibinfo {volume}
  {13}},\ \bibinfo {pages} {4972--4979} (\bibinfo {year} {2019})}\BibitemShut
  {NoStop}%
\bibitem [{\citenamefont {Gartner}, \citenamefont {Heil},\ and\ \citenamefont
  {Jayaraman}(2020)}]{Gartner2020}%
  \BibitemOpen
  \bibfield  {author} {\bibinfo {author} {\bibfnamefont {T.~E.}\ \bibnamefont
  {Gartner}}, \bibinfo {author} {\bibfnamefont {C.~M.}\ \bibnamefont {Heil}},\
  and\ \bibinfo {author} {\bibfnamefont {A.}~\bibnamefont {Jayaraman}},\
  }\bibfield  {title} {\enquote {\bibinfo {title} {Surface composition and
  ordering of binary nanoparticle mixtures in spherical confinement},}\ }\href
  {https://doi.org/10.1039/c9me00185a} {\bibfield  {journal} {\bibinfo
  {journal} {Molecular Systems Design {\&} Engineering}\ }\textbf {\bibinfo
  {volume} {5}},\ \bibinfo {pages} {864--875} (\bibinfo {year}
  {2020})}\BibitemShut {NoStop}%
\bibitem [{\citenamefont {Kim}\ \emph {et~al.}(2021{\natexlab{a}})\citenamefont
  {Kim}, \citenamefont {Shim}, \citenamefont {Jo},\ and\ \citenamefont
  {Wooh}}]{Kim2021}%
  \BibitemOpen
  \bibfield  {author} {\bibinfo {author} {\bibfnamefont {J.}~\bibnamefont
  {Kim}}, \bibinfo {author} {\bibfnamefont {W.}~\bibnamefont {Shim}}, \bibinfo
  {author} {\bibfnamefont {S.-M.}\ \bibnamefont {Jo}},\ and\ \bibinfo {author}
  {\bibfnamefont {S.}~\bibnamefont {Wooh}},\ }\bibfield  {title} {\enquote
  {\bibinfo {title} {Evaporation driven synthesis of supraparticles on liquid
  repellent surfaces},}\ }\href {https://doi.org/10.1016/j.jiec.2020.12.017}
  {\bibfield  {journal} {\bibinfo  {journal} {Journal of Industrial and
  Engineering Chemistry}\ }\textbf {\bibinfo {volume} {95}},\ \bibinfo {pages}
  {170--181} (\bibinfo {year} {2021}{\natexlab{a}})}\BibitemShut {NoStop}%
\bibitem [{\citenamefont {Raju}\ \emph {et~al.}(2021)\citenamefont {Raju},
  \citenamefont {Koshkina}, \citenamefont {Tan}, \citenamefont {Riedinger},
  \citenamefont {Landfester}, \citenamefont {Lohse},\ and\ \citenamefont
  {Zhang}}]{Raju2021}%
  \BibitemOpen
  \bibfield  {author} {\bibinfo {author} {\bibfnamefont {L.~T.}\ \bibnamefont
  {Raju}}, \bibinfo {author} {\bibfnamefont {O.}~\bibnamefont {Koshkina}},
  \bibinfo {author} {\bibfnamefont {H.}~\bibnamefont {Tan}}, \bibinfo {author}
  {\bibfnamefont {A.}~\bibnamefont {Riedinger}}, \bibinfo {author}
  {\bibfnamefont {K.}~\bibnamefont {Landfester}}, \bibinfo {author}
  {\bibfnamefont {D.}~\bibnamefont {Lohse}},\ and\ \bibinfo {author}
  {\bibfnamefont {X.}~\bibnamefont {Zhang}},\ }\bibfield  {title} {\enquote
  {\bibinfo {title} {Particle size determines the shape of supraparticles in
  self-lubricating ternary droplets},}\ }\href
  {https://doi.org/10.1021/acsnano.0c06814} {\bibfield  {journal} {\bibinfo
  {journal} {{ACS} Nano}\ }\textbf {\bibinfo {volume} {15}},\ \bibinfo {pages}
  {4256--4267} (\bibinfo {year} {2021})}\BibitemShut {NoStop}%
\bibitem [{\citenamefont {Wang}\ and\ \citenamefont {Keddie}(2009)}]{Wang2009}%
  \BibitemOpen
  \bibfield  {author} {\bibinfo {author} {\bibfnamefont {T.}~\bibnamefont
  {Wang}}\ and\ \bibinfo {author} {\bibfnamefont {J.~L.}\ \bibnamefont
  {Keddie}},\ }\bibfield  {title} {\enquote {\bibinfo {title} {Design and
  fabrication of colloidal polymer nanocomposites},}\ }\href
  {https://doi.org/10.1016/j.cis.2008.06.002} {\bibfield  {journal} {\bibinfo
  {journal} {Advances in Colloid and Interface Science}\ }\textbf {\bibinfo
  {volume} {147-148}},\ \bibinfo {pages} {319--332} (\bibinfo {year}
  {2009})}\BibitemShut {NoStop}%
\bibitem [{\citenamefont {Dong}\ \emph {et~al.}(2020)\citenamefont {Dong},
  \citenamefont {Busatto}, \citenamefont {Roth},\ and\ \citenamefont
  {Martin-Fabiani}}]{Dong2020}%
  \BibitemOpen
  \bibfield  {author} {\bibinfo {author} {\bibfnamefont {Y.}~\bibnamefont
  {Dong}}, \bibinfo {author} {\bibfnamefont {N.}~\bibnamefont {Busatto}},
  \bibinfo {author} {\bibfnamefont {P.~J.}\ \bibnamefont {Roth}},\ and\
  \bibinfo {author} {\bibfnamefont {I.}~\bibnamefont {Martin-Fabiani}},\
  }\bibfield  {title} {\enquote {\bibinfo {title} {Colloidal assembly of
  polydisperse particle blends during drying},}\ }\href
  {https://doi.org/10.1039/d0sm00785d} {\bibfield  {journal} {\bibinfo
  {journal} {Soft Matter}\ }\textbf {\bibinfo {volume} {16}},\ \bibinfo {pages}
  {8453--8461} (\bibinfo {year} {2020})}\BibitemShut {NoStop}%
\bibitem [{\citenamefont {Kim}\ \emph {et~al.}(2021{\natexlab{b}})\citenamefont
  {Kim}, \citenamefont {Qiang}, \citenamefont {Turner}, \citenamefont {Choi},\
  and\ \citenamefont {Lee}}]{Kim2021a}%
  \BibitemOpen
  \bibfield  {author} {\bibinfo {author} {\bibfnamefont {B.~Q.}\ \bibnamefont
  {Kim}}, \bibinfo {author} {\bibfnamefont {Y.}~\bibnamefont {Qiang}}, \bibinfo
  {author} {\bibfnamefont {K.~T.}\ \bibnamefont {Turner}}, \bibinfo {author}
  {\bibfnamefont {S.~Q.}\ \bibnamefont {Choi}},\ and\ \bibinfo {author}
  {\bibfnamefont {D.}~\bibnamefont {Lee}},\ }\bibfield  {title} {\enquote
  {\bibinfo {title} {Heterostructured polymer-infiltrated nanoparticle films
  with cavities via capillary rise infiltration},}\ }\href
  {https://doi.org/https://doi.org/10.1002/admi.202001421} {\bibfield
  {journal} {\bibinfo  {journal} {Advanced Materials Interfaces}\ }\textbf
  {\bibinfo {volume} {8}},\ \bibinfo {pages} {2001421} (\bibinfo {year}
  {2021}{\natexlab{b}})}\BibitemShut {NoStop}%
\bibitem [{\citenamefont {Choi}\ \emph {et~al.}(2010)\citenamefont {Choi},
  \citenamefont {Stassi}, \citenamefont {Pisano},\ and\ \citenamefont
  {Zohdi}}]{Choi2010}%
  \BibitemOpen
  \bibfield  {author} {\bibinfo {author} {\bibfnamefont {S.}~\bibnamefont
  {Choi}}, \bibinfo {author} {\bibfnamefont {S.}~\bibnamefont {Stassi}},
  \bibinfo {author} {\bibfnamefont {A.~P.}\ \bibnamefont {Pisano}},\ and\
  \bibinfo {author} {\bibfnamefont {T.~I.}\ \bibnamefont {Zohdi}},\ }\bibfield
  {title} {\enquote {\bibinfo {title} {Coffee-ring effect-based three
  dimensional patterning of micro/nanoparticle assembly with a single
  droplet},}\ }\href {https://doi.org/10.1021/la101110t} {\bibfield  {journal}
  {\bibinfo  {journal} {Langmuir}\ }\textbf {\bibinfo {volume} {26}},\ \bibinfo
  {pages} {11690--11698} (\bibinfo {year} {2010})}\BibitemShut {NoStop}%
\bibitem [{\citenamefont {Nunes}\ \emph {et~al.}(2020)\citenamefont {Nunes},
  \citenamefont {Velu}, \citenamefont {Kasianiuk}, \citenamefont {Kasyanyuk},
  \citenamefont {Callegari}, \citenamefont {Volpe}, \citenamefont {da~Gama},
  \citenamefont {Volpe},\ and\ \citenamefont {Ara{\'{u}}jo}}]{Nunes2020}%
  \BibitemOpen
  \bibfield  {author} {\bibinfo {author} {\bibfnamefont {A.~S.}\ \bibnamefont
  {Nunes}}, \bibinfo {author} {\bibfnamefont {S.~K.~P.}\ \bibnamefont {Velu}},
  \bibinfo {author} {\bibfnamefont {I.}~\bibnamefont {Kasianiuk}}, \bibinfo
  {author} {\bibfnamefont {D.}~\bibnamefont {Kasyanyuk}}, \bibinfo {author}
  {\bibfnamefont {A.}~\bibnamefont {Callegari}}, \bibinfo {author}
  {\bibfnamefont {G.}~\bibnamefont {Volpe}}, \bibinfo {author} {\bibfnamefont
  {M.~M.~T.}\ \bibnamefont {da~Gama}}, \bibinfo {author} {\bibfnamefont
  {G.}~\bibnamefont {Volpe}},\ and\ \bibinfo {author} {\bibfnamefont
  {N.~A.~M.}\ \bibnamefont {Ara{\'{u}}jo}},\ }\bibfield  {title} {\enquote
  {\bibinfo {title} {Ordering of binary colloidal crystals by random
  potentials},}\ }\href {https://doi.org/10.1039/d0sm00208a} {\bibfield
  {journal} {\bibinfo  {journal} {Soft Matter}\ }\textbf {\bibinfo {volume}
  {16}},\ \bibinfo {pages} {4267--4273} (\bibinfo {year} {2020})}\BibitemShut
  {NoStop}%
\bibitem [{\citenamefont {Nozawa}\ \emph {et~al.}(2022)\citenamefont {Nozawa},
  \citenamefont {Uda}, \citenamefont {Toyotama}, \citenamefont {Yamanaka},
  \citenamefont {Niinomi},\ and\ \citenamefont {Okada}}]{Nozawa2022}%
  \BibitemOpen
  \bibfield  {author} {\bibinfo {author} {\bibfnamefont {J.}~\bibnamefont
  {Nozawa}}, \bibinfo {author} {\bibfnamefont {S.}~\bibnamefont {Uda}},
  \bibinfo {author} {\bibfnamefont {A.}~\bibnamefont {Toyotama}}, \bibinfo
  {author} {\bibfnamefont {J.}~\bibnamefont {Yamanaka}}, \bibinfo {author}
  {\bibfnamefont {H.}~\bibnamefont {Niinomi}},\ and\ \bibinfo {author}
  {\bibfnamefont {J.}~\bibnamefont {Okada}},\ }\bibfield  {title} {\enquote
  {\bibinfo {title} {Heteroepitaxial fabrication of binary colloidal crystals
  by a balance of interparticle interaction and lattice spacing},}\ }\href
  {https://doi.org/10.1016/j.jcis.2021.10.041} {\bibfield  {journal} {\bibinfo
  {journal} {Journal of Colloid and Interface Science}\ }\textbf {\bibinfo
  {volume} {608}},\ \bibinfo {pages} {873--881} (\bibinfo {year}
  {2022})}\BibitemShut {NoStop}%
\bibitem [{\citenamefont {Das}\ \emph {et~al.}(2018)\citenamefont {Das},
  \citenamefont {Duraia}, \citenamefont {Velev}, \citenamefont {Amiri},\ and\
  \citenamefont {Beall}}]{Das2018}%
  \BibitemOpen
  \bibfield  {author} {\bibinfo {author} {\bibfnamefont {S.}~\bibnamefont
  {Das}}, \bibinfo {author} {\bibfnamefont {E.-S.~M.}\ \bibnamefont {Duraia}},
  \bibinfo {author} {\bibfnamefont {O.~D.}\ \bibnamefont {Velev}}, \bibinfo
  {author} {\bibfnamefont {M.~D.}\ \bibnamefont {Amiri}},\ and\ \bibinfo
  {author} {\bibfnamefont {G.~W.}\ \bibnamefont {Beall}},\ }\bibfield  {title}
  {\enquote {\bibinfo {title} {Formation of periodic size-segregated stripe
  pattern via directed self-assembly of binary colloids and its mechanism},}\
  }\href {https://doi.org/10.1016/j.apsusc.2017.11.142} {\bibfield  {journal}
  {\bibinfo  {journal} {Applied Surface Science}\ }\textbf {\bibinfo {volume}
  {435}},\ \bibinfo {pages} {512--520} (\bibinfo {year} {2018})}\BibitemShut
  {NoStop}%
\bibitem [{\citenamefont {Li}\ and\ \citenamefont {Garno}(2009)}]{Li2009969}%
  \BibitemOpen
  \bibfield  {author} {\bibinfo {author} {\bibfnamefont {J.-R.}\ \bibnamefont
  {Li}}\ and\ \bibinfo {author} {\bibfnamefont {J.~C.}\ \bibnamefont {Garno}},\
  }\bibfield  {title} {\enquote {\bibinfo {title} {Nanostructures of
  octadecyltrisiloxane self-assembled monolayers produced on {A}u(111) using
  particle lithography},}\ }\href {https://doi.org/10.1021/am900118x}
  {\bibfield  {journal} {\bibinfo  {journal} {{ACS} Applied Materials {\&}
  Interfaces}\ }\textbf {\bibinfo {volume} {1}},\ \bibinfo {pages} {969--976}
  (\bibinfo {year} {2009})}\BibitemShut {NoStop}%
\bibitem [{\citenamefont {Li}\ \emph {et~al.}(2009)\citenamefont {Li},
  \citenamefont {Lusker}, \citenamefont {Yu},\ and\ \citenamefont
  {Garno}}]{Li2009}%
  \BibitemOpen
  \bibfield  {author} {\bibinfo {author} {\bibfnamefont {J.-R.}\ \bibnamefont
  {Li}}, \bibinfo {author} {\bibfnamefont {K.~L.}\ \bibnamefont {Lusker}},
  \bibinfo {author} {\bibfnamefont {J.-J.}\ \bibnamefont {Yu}},\ and\ \bibinfo
  {author} {\bibfnamefont {J.~C.}\ \bibnamefont {Garno}},\ }\bibfield  {title}
  {\enquote {\bibinfo {title} {Engineering the spatial selectivity of surfaces
  at the nanoscale using particle lithography combined with vapor deposition of
  organosilanes},}\ }\href {https://doi.org/10.1021/nn9004796} {\bibfield
  {journal} {\bibinfo  {journal} {{ACS} Nano}\ }\textbf {\bibinfo {volume}
  {3}},\ \bibinfo {pages} {2023--2035} (\bibinfo {year} {2009})}\BibitemShut
  {NoStop}%
\bibitem [{\citenamefont {Chen}\ \emph {et~al.}(2009)\citenamefont {Chen},
  \citenamefont {Liao}, \citenamefont {Chen}, \citenamefont {Yang},
  \citenamefont {Wark}, \citenamefont {Son}, \citenamefont {Batteas},\ and\
  \citenamefont {Cremer}}]{Chen2009}%
  \BibitemOpen
  \bibfield  {author} {\bibinfo {author} {\bibfnamefont {J.}~\bibnamefont
  {Chen}}, \bibinfo {author} {\bibfnamefont {W.-S.}\ \bibnamefont {Liao}},
  \bibinfo {author} {\bibfnamefont {X.}~\bibnamefont {Chen}}, \bibinfo {author}
  {\bibfnamefont {T.}~\bibnamefont {Yang}}, \bibinfo {author} {\bibfnamefont
  {S.~E.}\ \bibnamefont {Wark}}, \bibinfo {author} {\bibfnamefont {D.~H.}\
  \bibnamefont {Son}}, \bibinfo {author} {\bibfnamefont {J.~D.}\ \bibnamefont
  {Batteas}},\ and\ \bibinfo {author} {\bibfnamefont {P.~S.}\ \bibnamefont
  {Cremer}},\ }\bibfield  {title} {\enquote {\bibinfo {title}
  {Evaporation-induced assembly of quantum dots into nanorings},}\ }\href
  {https://doi.org/10.1021/nn800568t} {\bibfield  {journal} {\bibinfo
  {journal} {{ACS} Nano}\ }\textbf {\bibinfo {volume} {3}},\ \bibinfo {pages}
  {173--180} (\bibinfo {year} {2009})}\BibitemShut {NoStop}%
\bibitem [{\citenamefont {Utgenannt}\ \emph {et~al.}(2016)\citenamefont
  {Utgenannt}, \citenamefont {Maspero}, \citenamefont {Fortini}, \citenamefont
  {Turner}, \citenamefont {Florescu}, \citenamefont {Jeynes}, \citenamefont
  {Kanaras}, \citenamefont {Muskens}, \citenamefont {Sear},\ and\ \citenamefont
  {Keddie}}]{Utgenannt2016}%
  \BibitemOpen
  \bibfield  {author} {\bibinfo {author} {\bibfnamefont {A.}~\bibnamefont
  {Utgenannt}}, \bibinfo {author} {\bibfnamefont {R.}~\bibnamefont {Maspero}},
  \bibinfo {author} {\bibfnamefont {A.}~\bibnamefont {Fortini}}, \bibinfo
  {author} {\bibfnamefont {R.}~\bibnamefont {Turner}}, \bibinfo {author}
  {\bibfnamefont {M.}~\bibnamefont {Florescu}}, \bibinfo {author}
  {\bibfnamefont {C.}~\bibnamefont {Jeynes}}, \bibinfo {author} {\bibfnamefont
  {A.~G.}\ \bibnamefont {Kanaras}}, \bibinfo {author} {\bibfnamefont {O.~L.}\
  \bibnamefont {Muskens}}, \bibinfo {author} {\bibfnamefont {R.~P.}\
  \bibnamefont {Sear}},\ and\ \bibinfo {author} {\bibfnamefont {J.~L.}\
  \bibnamefont {Keddie}},\ }\bibfield  {title} {\enquote {\bibinfo {title}
  {Fast assembly of gold nanoparticles in large-area 2{D} nanogrids using a
  one-step, near-infrared radiation-assisted evaporation process},}\ }\href
  {https://doi.org/10.1021/acsnano.5b06886} {\bibfield  {journal} {\bibinfo
  {journal} {{ACS} Nano}\ }\textbf {\bibinfo {volume} {10}},\ \bibinfo {pages}
  {2232--2242} (\bibinfo {year} {2016})}\BibitemShut {NoStop}%
\bibitem [{\citenamefont {Al-Milaji}\ and\ \citenamefont
  {Zhao}(2019)}]{AlMilaji2019Langmuir}%
  \BibitemOpen
  \bibfield  {author} {\bibinfo {author} {\bibfnamefont {K.~N.}\ \bibnamefont
  {Al-Milaji}}\ and\ \bibinfo {author} {\bibfnamefont {H.}~\bibnamefont
  {Zhao}},\ }\bibfield  {title} {\enquote {\bibinfo {title} {Probing the
  colloidal particle dynamics in drying sessile droplets},}\ }\href
  {https://doi.org/10.1021/acs.langmuir.8b03406} {\bibfield  {journal}
  {\bibinfo  {journal} {Langmuir}\ }\textbf {\bibinfo {volume} {35}},\ \bibinfo
  {pages} {2209--2220} (\bibinfo {year} {2019})}\BibitemShut {NoStop}%
\bibitem [{\citenamefont {Detrich}\ \emph {et~al.}(2009)\citenamefont
  {Detrich}, \citenamefont {De{\'{a}}k}, \citenamefont {Hild}, \citenamefont
  {Kov{\'{a}}cs},\ and\ \citenamefont {H{\'{o}}rv{\"{o}}lgyi}}]{Detrich2009}%
  \BibitemOpen
  \bibfield  {author} {\bibinfo {author} {\bibfnamefont {{\'{A}}.}~\bibnamefont
  {Detrich}}, \bibinfo {author} {\bibfnamefont {A.}~\bibnamefont {De{\'{a}}k}},
  \bibinfo {author} {\bibfnamefont {E.}~\bibnamefont {Hild}}, \bibinfo {author}
  {\bibfnamefont {A.~L.}\ \bibnamefont {Kov{\'{a}}cs}},\ and\ \bibinfo {author}
  {\bibfnamefont {Z.}~\bibnamefont {H{\'{o}}rv{\"{o}}lgyi}},\ }\bibfield
  {title} {\enquote {\bibinfo {title} {Langmuir and {L}angmuir-{B}lodgett films
  of bidisperse silica nanoparticles},}\ }\href
  {https://doi.org/10.1021/la9027207} {\bibfield  {journal} {\bibinfo
  {journal} {Langmuir}\ }\textbf {\bibinfo {volume} {26}},\ \bibinfo {pages}
  {2694--2699} (\bibinfo {year} {2009})}\BibitemShut {NoStop}%
\bibitem [{\citenamefont {Vogel}\ \emph {et~al.}(2011)\citenamefont {Vogel},
  \citenamefont {de~Viguerie}, \citenamefont {Jonas}, \citenamefont {Weiss},\
  and\ \citenamefont {Landfester}}]{Vogel2011}%
  \BibitemOpen
  \bibfield  {author} {\bibinfo {author} {\bibfnamefont {N.}~\bibnamefont
  {Vogel}}, \bibinfo {author} {\bibfnamefont {L.}~\bibnamefont {de~Viguerie}},
  \bibinfo {author} {\bibfnamefont {U.}~\bibnamefont {Jonas}}, \bibinfo
  {author} {\bibfnamefont {C.~K.}\ \bibnamefont {Weiss}},\ and\ \bibinfo
  {author} {\bibfnamefont {K.}~\bibnamefont {Landfester}},\ }\bibfield  {title}
  {\enquote {\bibinfo {title} {Wafer-scale fabrication of ordered binary
  colloidal monolayers with adjustable stoichiometries},}\ }\href
  {https://doi.org/10.1002/adfm.201100414} {\bibfield  {journal} {\bibinfo
  {journal} {Advanced Functional Materials}\ }\textbf {\bibinfo {volume}
  {21}},\ \bibinfo {pages} {3064--3073} (\bibinfo {year} {2011})}\BibitemShut
  {NoStop}%
\bibitem [{\citenamefont {Sharma}\ \emph {et~al.}(2009)\citenamefont {Sharma},
  \citenamefont {Yan}, \citenamefont {Wong}, \citenamefont {Carter},\ and\
  \citenamefont {Chiang}}]{Sharma2009}%
  \BibitemOpen
  \bibfield  {author} {\bibinfo {author} {\bibfnamefont {V.}~\bibnamefont
  {Sharma}}, \bibinfo {author} {\bibfnamefont {Q.}~\bibnamefont {Yan}},
  \bibinfo {author} {\bibfnamefont {C.~C.}\ \bibnamefont {Wong}}, \bibinfo
  {author} {\bibfnamefont {W.~C.}\ \bibnamefont {Carter}},\ and\ \bibinfo
  {author} {\bibfnamefont {Y.-M.}\ \bibnamefont {Chiang}},\ }\bibfield  {title}
  {\enquote {\bibinfo {title} {Controlled and rapid ordering of oppositely
  charged colloidal particles},}\ }\href
  {https://doi.org/10.1016/j.jcis.2009.01.047} {\bibfield  {journal} {\bibinfo
  {journal} {Journal of Colloid and Interface Science}\ }\textbf {\bibinfo
  {volume} {333}},\ \bibinfo {pages} {230--236} (\bibinfo {year}
  {2009})}\BibitemShut {NoStop}%
\bibitem [{\citenamefont {Lotito}\ and\ \citenamefont
  {Zambelli}(2016)}]{Lotito2016}%
  \BibitemOpen
  \bibfield  {author} {\bibinfo {author} {\bibfnamefont {V.}~\bibnamefont
  {Lotito}}\ and\ \bibinfo {author} {\bibfnamefont {T.}~\bibnamefont
  {Zambelli}},\ }\bibfield  {title} {\enquote {\bibinfo {title} {Self-assembly
  of single-sized and binary colloidal particles at air/water interface by
  surface confinement and water discharge},}\ }\href
  {https://doi.org/10.1021/acs.langmuir.6b02157} {\bibfield  {journal}
  {\bibinfo  {journal} {Langmuir}\ }\textbf {\bibinfo {volume} {32}},\ \bibinfo
  {pages} {9582--9590} (\bibinfo {year} {2016})}\BibitemShut {NoStop}%
\bibitem [{\citenamefont {Inoue}\ and\ \citenamefont
  {Inasawa}(2020)}]{Inoue2020}%
  \BibitemOpen
  \bibfield  {author} {\bibinfo {author} {\bibfnamefont {K.}~\bibnamefont
  {Inoue}}\ and\ \bibinfo {author} {\bibfnamefont {S.}~\bibnamefont
  {Inasawa}},\ }\bibfield  {title} {\enquote {\bibinfo {title} {Positive and
  negative birefringence in packed films of binary spherical colloidal
  particles},}\ }\href {https://doi.org/10.1039/c9ra09704j} {\bibfield
  {journal} {\bibinfo  {journal} {{RSC} Advances}\ }\textbf {\bibinfo {volume}
  {10}},\ \bibinfo {pages} {2566--2574} (\bibinfo {year} {2020})}\BibitemShut
  {NoStop}%
\bibitem [{\citenamefont {Guo}\ \emph {et~al.}(2017)\citenamefont {Guo},
  \citenamefont {Li}, \citenamefont {Zheng}, \citenamefont {Li}, \citenamefont
  {Chen}, \citenamefont {Li}, \citenamefont {Yang}, \citenamefont {Li},\ and\
  \citenamefont {Song}}]{Guo2017}%
  \BibitemOpen
  \bibfield  {author} {\bibinfo {author} {\bibfnamefont {D.}~\bibnamefont
  {Guo}}, \bibinfo {author} {\bibfnamefont {Y.}~\bibnamefont {Li}}, \bibinfo
  {author} {\bibfnamefont {X.}~\bibnamefont {Zheng}}, \bibinfo {author}
  {\bibfnamefont {F.}~\bibnamefont {Li}}, \bibinfo {author} {\bibfnamefont
  {S.}~\bibnamefont {Chen}}, \bibinfo {author} {\bibfnamefont {M.}~\bibnamefont
  {Li}}, \bibinfo {author} {\bibfnamefont {Q.}~\bibnamefont {Yang}}, \bibinfo
  {author} {\bibfnamefont {H.}~\bibnamefont {Li}},\ and\ \bibinfo {author}
  {\bibfnamefont {Y.}~\bibnamefont {Song}},\ }\bibfield  {title} {\enquote
  {\bibinfo {title} {Programmed coassembly of one-dimensional binary
  superstructures by liquid soft confinement},}\ }\href
  {https://doi.org/10.1021/jacs.7b09738} {\bibfield  {journal} {\bibinfo
  {journal} {Journal of the American Chemical Society}\ }\textbf {\bibinfo
  {volume} {140}},\ \bibinfo {pages} {18--21} (\bibinfo {year}
  {2017})}\BibitemShut {NoStop}%
\bibitem [{\citenamefont {Schulz}\ \emph {et~al.}(2020)\citenamefont {Schulz},
  \citenamefont {Smith}, \citenamefont {Sear}, \citenamefont {Brinkhuis},\ and\
  \citenamefont {Keddie}}]{Schulz2020}%
  \BibitemOpen
  \bibfield  {author} {\bibinfo {author} {\bibfnamefont {M.}~\bibnamefont
  {Schulz}}, \bibinfo {author} {\bibfnamefont {R.~W.}\ \bibnamefont {Smith}},
  \bibinfo {author} {\bibfnamefont {R.~P.}\ \bibnamefont {Sear}}, \bibinfo
  {author} {\bibfnamefont {R.}~\bibnamefont {Brinkhuis}},\ and\ \bibinfo
  {author} {\bibfnamefont {J.~L.}\ \bibnamefont {Keddie}},\ }\bibfield  {title}
  {\enquote {\bibinfo {title} {Diffusiophoresis-driven stratification of
  polymers in colloidal films},}\ }\href
  {https://doi.org/10.1021/acsmacrolett.0c00363} {\bibfield  {journal}
  {\bibinfo  {journal} {{ACS} Macro Letters}\ }\textbf {\bibinfo {volume}
  {9}},\ \bibinfo {pages} {1286--1291} (\bibinfo {year} {2020})}\BibitemShut
  {NoStop}%
\bibitem [{\citenamefont {Jeong}, \citenamefont {Lee},\ and\ \citenamefont
  {Ahn}(2021)}]{Jeong2021}%
  \BibitemOpen
  \bibfield  {author} {\bibinfo {author} {\bibfnamefont {J.~H.}\ \bibnamefont
  {Jeong}}, \bibinfo {author} {\bibfnamefont {Y.~K.}\ \bibnamefont {Lee}},\
  and\ \bibinfo {author} {\bibfnamefont {K.~H.}\ \bibnamefont {Ahn}},\
  }\bibfield  {title} {\enquote {\bibinfo {title} {Stratification mechanism in
  the bidisperse colloidal film drying process: evolution and decomposition of
  normal stress correlated with microstructure},}\ }\href
  {https://doi.org/10.1021/acs.langmuir.1c02455} {\bibfield  {journal}
  {\bibinfo  {journal} {Langmuir}\ } (\bibinfo {year} {2021}),\
  10.1021/acs.langmuir.1c02455}\BibitemShut {NoStop}%
\bibitem [{\citenamefont {Samanta}\ and\ \citenamefont
  {Bordes}(2020)}]{Samanta2020}%
  \BibitemOpen
  \bibfield  {author} {\bibinfo {author} {\bibfnamefont {A.}~\bibnamefont
  {Samanta}}\ and\ \bibinfo {author} {\bibfnamefont {R.}~\bibnamefont
  {Bordes}},\ }\bibfield  {title} {\enquote {\bibinfo {title} {On the effect of
  particle surface chemistry in film stratification and morphology
  regulation},}\ }\href {https://doi.org/10.1039/d0sm00317d} {\bibfield
  {journal} {\bibinfo  {journal} {Soft Matter}\ }\textbf {\bibinfo {volume}
  {16}},\ \bibinfo {pages} {6371--6378} (\bibinfo {year} {2020})}\BibitemShut
  {NoStop}%
\bibitem [{\citenamefont {Atmuri}, \citenamefont {Bhatia},\ and\ \citenamefont
  {Routh}(2012)}]{Atmuri2012}%
  \BibitemOpen
  \bibfield  {author} {\bibinfo {author} {\bibfnamefont {A.~K.}\ \bibnamefont
  {Atmuri}}, \bibinfo {author} {\bibfnamefont {S.~R.}\ \bibnamefont {Bhatia}},\
  and\ \bibinfo {author} {\bibfnamefont {A.~F.}\ \bibnamefont {Routh}},\
  }\bibfield  {title} {\enquote {\bibinfo {title} {Autostratification in drying
  colloidal dispersions: effect of particle interactions},}\ }\href
  {https://doi.org/10.1021/la2039762} {\bibfield  {journal} {\bibinfo
  {journal} {Langmuir}\ }\textbf {\bibinfo {volume} {28}},\ \bibinfo {pages}
  {2652--2658} (\bibinfo {year} {2012})}\BibitemShut {NoStop}%
\bibitem [{\citenamefont {He}\ \emph {et~al.}(2021)\citenamefont {He},
  \citenamefont {Martin-Fabiani}, \citenamefont {Roth}, \citenamefont
  {T{\'{o}}th},\ and\ \citenamefont {Archer}}]{He2021}%
  \BibitemOpen
  \bibfield  {author} {\bibinfo {author} {\bibfnamefont {B.}~\bibnamefont
  {He}}, \bibinfo {author} {\bibfnamefont {I.}~\bibnamefont {Martin-Fabiani}},
  \bibinfo {author} {\bibfnamefont {R.}~\bibnamefont {Roth}}, \bibinfo {author}
  {\bibfnamefont {G.~I.}\ \bibnamefont {T{\'{o}}th}},\ and\ \bibinfo {author}
  {\bibfnamefont {A.~J.}\ \bibnamefont {Archer}},\ }\bibfield  {title}
  {\enquote {\bibinfo {title} {Dynamical density functional theory for the
  drying and stratification of binary colloidal dispersions},}\ }\href
  {https://doi.org/10.1021/acs.langmuir.0c02825} {\bibfield  {journal}
  {\bibinfo  {journal} {Langmuir}\ }\textbf {\bibinfo {volume} {37}},\ \bibinfo
  {pages} {1399--1409} (\bibinfo {year} {2021})}\BibitemShut {NoStop}%
\bibitem [{\citenamefont {Tang}, \citenamefont {Grest},\ and\ \citenamefont
  {Cheng}(2019)}]{Tang2019}%
  \BibitemOpen
  \bibfield  {author} {\bibinfo {author} {\bibfnamefont {Y.}~\bibnamefont
  {Tang}}, \bibinfo {author} {\bibfnamefont {G.~S.}\ \bibnamefont {Grest}},\
  and\ \bibinfo {author} {\bibfnamefont {S.}~\bibnamefont {Cheng}},\ }\bibfield
   {title} {\enquote {\bibinfo {title} {Stratification of drying particle
  suspensions: Comparison of implicit and explicit solvent simulations},}\
  }\href {https://doi.org/10.1063/1.5066035} {\bibfield  {journal} {\bibinfo
  {journal} {The Journal of Chemical Physics}\ }\textbf {\bibinfo {volume}
  {150}},\ \bibinfo {pages} {224901} (\bibinfo {year} {2019})}\BibitemShut
  {NoStop}%
\bibitem [{\citenamefont {Zhou}, \citenamefont {Jiang},\ and\ \citenamefont
  {Doi}(2017)}]{Zhou2017PRL}%
  \BibitemOpen
  \bibfield  {author} {\bibinfo {author} {\bibfnamefont {J.}~\bibnamefont
  {Zhou}}, \bibinfo {author} {\bibfnamefont {Y.}~\bibnamefont {Jiang}},\ and\
  \bibinfo {author} {\bibfnamefont {M.}~\bibnamefont {Doi}},\ }\bibfield
  {title} {\enquote {\bibinfo {title} {Cross interaction drives stratification
  in drying film of binary colloidal mixtures},}\ }\href
  {https://doi.org/10.1103/physrevlett.118.108002} {\bibfield  {journal}
  {\bibinfo  {journal} {Physical Review Letters}\ }\textbf {\bibinfo {volume}
  {118}},\ \bibinfo {pages} {108002} (\bibinfo {year} {2017})}\BibitemShut
  {NoStop}%
\bibitem [{\citenamefont {Zhou}\ \emph {et~al.}(2017)\citenamefont {Zhou},
  \citenamefont {Man}, \citenamefont {Jiang},\ and\ \citenamefont
  {Doi}}]{Zhou2017}%
  \BibitemOpen
  \bibfield  {author} {\bibinfo {author} {\bibfnamefont {J.}~\bibnamefont
  {Zhou}}, \bibinfo {author} {\bibfnamefont {X.}~\bibnamefont {Man}}, \bibinfo
  {author} {\bibfnamefont {Y.}~\bibnamefont {Jiang}},\ and\ \bibinfo {author}
  {\bibfnamefont {M.}~\bibnamefont {Doi}},\ }\bibfield  {title} {\enquote
  {\bibinfo {title} {Structure formation in soft-matter solutions induced by
  solvent evaporation},}\ }\href {https://doi.org/10.1002/adma.201703769}
  {\bibfield  {journal} {\bibinfo  {journal} {Advanced Materials}\ }\textbf
  {\bibinfo {volume} {29}},\ \bibinfo {pages} {1703769} (\bibinfo {year}
  {2017})}\BibitemShut {NoStop}%
\bibitem [{\citenamefont {Schulz}\ and\ \citenamefont
  {Keddie}(2018)}]{Schulz2018}%
  \BibitemOpen
  \bibfield  {author} {\bibinfo {author} {\bibfnamefont {M.}~\bibnamefont
  {Schulz}}\ and\ \bibinfo {author} {\bibfnamefont {J.~L.}\ \bibnamefont
  {Keddie}},\ }\bibfield  {title} {\enquote {\bibinfo {title} {A critical and
  quantitative review of the stratification of particles during the drying of
  colloidal films},}\ }\href {https://doi.org/10.1039/c8sm01025k} {\bibfield
  {journal} {\bibinfo  {journal} {Soft Matter}\ }\textbf {\bibinfo {volume}
  {14}},\ \bibinfo {pages} {6181--6197} (\bibinfo {year} {2018})}\BibitemShut
  {NoStop}%
\bibitem [{\citenamefont {Yu}\ \emph {et~al.}(2021)\citenamefont {Yu},
  \citenamefont {Floch-Fou{\'{e}}r{\'{e}}}, \citenamefont {Pauchard},
  \citenamefont {Boissel}, \citenamefont {Fu}, \citenamefont {Chen},
  \citenamefont {Saint-Jalmes}, \citenamefont {Jeantet},\ and\ \citenamefont
  {Lanotte}}]{Yu2021}%
  \BibitemOpen
  \bibfield  {author} {\bibinfo {author} {\bibfnamefont {M.}~\bibnamefont
  {Yu}}, \bibinfo {author} {\bibfnamefont {C.~L.}\ \bibnamefont
  {Floch-Fou{\'{e}}r{\'{e}}}}, \bibinfo {author} {\bibfnamefont
  {L.}~\bibnamefont {Pauchard}}, \bibinfo {author} {\bibfnamefont
  {F.}~\bibnamefont {Boissel}}, \bibinfo {author} {\bibfnamefont
  {N.}~\bibnamefont {Fu}}, \bibinfo {author} {\bibfnamefont {X.~D.}\
  \bibnamefont {Chen}}, \bibinfo {author} {\bibfnamefont {A.}~\bibnamefont
  {Saint-Jalmes}}, \bibinfo {author} {\bibfnamefont {R.}~\bibnamefont
  {Jeantet}},\ and\ \bibinfo {author} {\bibfnamefont {L.}~\bibnamefont
  {Lanotte}},\ }\bibfield  {title} {\enquote {\bibinfo {title} {Skin layer
  stratification in drying droplets of dairy colloids},}\ }\href
  {https://doi.org/10.1016/j.colsurfa.2021.126560} {\bibfield  {journal}
  {\bibinfo  {journal} {Colloids and Surfaces A: Physicochemical and
  Engineering Aspects}\ }\textbf {\bibinfo {volume} {620}},\ \bibinfo {pages}
  {126560} (\bibinfo {year} {2021})}\BibitemShut {NoStop}%
\bibitem [{\citenamefont {Kumnorkaew}\ and\ \citenamefont
  {Gilchrist}(2009)}]{Kumnorkaew2009}%
  \BibitemOpen
  \bibfield  {author} {\bibinfo {author} {\bibfnamefont {P.}~\bibnamefont
  {Kumnorkaew}}\ and\ \bibinfo {author} {\bibfnamefont {J.~F.}\ \bibnamefont
  {Gilchrist}},\ }\bibfield  {title} {\enquote {\bibinfo {title} {Effect of
  nanoparticle concentration on the convective deposition of binary
  suspensions},}\ }\href {https://doi.org/10.1021/la804209m} {\bibfield
  {journal} {\bibinfo  {journal} {Langmuir}\ }\textbf {\bibinfo {volume}
  {25}},\ \bibinfo {pages} {6070--6075} (\bibinfo {year} {2009})}\BibitemShut
  {NoStop}%
\bibitem [{\citenamefont {Chhasatia}\ and\ \citenamefont
  {Sun}(2011)}]{Chhasatia2011}%
  \BibitemOpen
  \bibfield  {author} {\bibinfo {author} {\bibfnamefont {V.~H.}\ \bibnamefont
  {Chhasatia}}\ and\ \bibinfo {author} {\bibfnamefont {Y.}~\bibnamefont
  {Sun}},\ }\bibfield  {title} {\enquote {\bibinfo {title} {Interaction of
  bi-dispersed particles with contact line in an evaporating colloidal drop},}\
  }\href {https://doi.org/10.1039/c1sm06393f} {\bibfield  {journal} {\bibinfo
  {journal} {Soft Matter}\ }\textbf {\bibinfo {volume} {7}},\ \bibinfo {pages}
  {10135} (\bibinfo {year} {2011})}\BibitemShut {NoStop}%
\bibitem [{\citenamefont {Das}\ \emph {et~al.}(2012)\citenamefont {Das},
  \citenamefont {Waghmare}, \citenamefont {Fan}, \citenamefont {Gunda},
  \citenamefont {Roy},\ and\ \citenamefont {Mitra}}]{Das2012}%
  \BibitemOpen
  \bibfield  {author} {\bibinfo {author} {\bibfnamefont {S.}~\bibnamefont
  {Das}}, \bibinfo {author} {\bibfnamefont {P.~R.}\ \bibnamefont {Waghmare}},
  \bibinfo {author} {\bibfnamefont {M.}~\bibnamefont {Fan}}, \bibinfo {author}
  {\bibfnamefont {N.~S.~K.}\ \bibnamefont {Gunda}}, \bibinfo {author}
  {\bibfnamefont {S.~S.}\ \bibnamefont {Roy}},\ and\ \bibinfo {author}
  {\bibfnamefont {S.~K.}\ \bibnamefont {Mitra}},\ }\bibfield  {title} {\enquote
  {\bibinfo {title} {Dynamics of liquid droplets in an evaporating drop: liquid
  droplet {\textquotedblleft}coffee stain{\textquotedblright} effect},}\ }\href
  {https://doi.org/10.1039/c2ra20743e} {\bibfield  {journal} {\bibinfo
  {journal} {{RSC} Advances}\ }\textbf {\bibinfo {volume} {2}},\ \bibinfo
  {pages} {8390} (\bibinfo {year} {2012})}\BibitemShut {NoStop}%
\bibitem [{\citenamefont {Devlin}, \citenamefont {Loehr},\ and\ \citenamefont
  {Harris}(2015)}]{Devlin2015}%
  \BibitemOpen
  \bibfield  {author} {\bibinfo {author} {\bibfnamefont {N.~R.}\ \bibnamefont
  {Devlin}}, \bibinfo {author} {\bibfnamefont {K.}~\bibnamefont {Loehr}},\ and\
  \bibinfo {author} {\bibfnamefont {M.~T.}\ \bibnamefont {Harris}},\ }\bibfield
   {title} {\enquote {\bibinfo {title} {The separation of two different sized
  particles in an evaporating droplet},}\ }\href
  {https://doi.org/10.1002/aic.14977} {\bibfield  {journal} {\bibinfo
  {journal} {{AIChE} Journal}\ }\textbf {\bibinfo {volume} {61}},\ \bibinfo
  {pages} {3547--3556} (\bibinfo {year} {2015})}\BibitemShut {NoStop}%
\bibitem [{\citenamefont {Yi}, \citenamefont {Jeong},\ and\ \citenamefont
  {Park}(2018)}]{Yi2018}%
  \BibitemOpen
  \bibfield  {author} {\bibinfo {author} {\bibfnamefont {J.}~\bibnamefont
  {Yi}}, \bibinfo {author} {\bibfnamefont {H.}~\bibnamefont {Jeong}},\ and\
  \bibinfo {author} {\bibfnamefont {J.}~\bibnamefont {Park}},\ }\bibfield
  {title} {\enquote {\bibinfo {title} {Modulation of nanoparticle separation by
  initial contact angle in coffee ring effect},}\ }\href
  {https://doi.org/10.1186/s40486-018-0079-9} {\bibfield  {journal} {\bibinfo
  {journal} {Micro and Nano Systems Letters}\ }\textbf {\bibinfo {volume} {6}}
  (\bibinfo {year} {2018}),\ 10.1186/s40486-018-0079-9}\BibitemShut {NoStop}%
\bibitem [{\citenamefont {Singh}\ \emph
  {et~al.}(2011{\natexlab{a}})\citenamefont {Singh}, \citenamefont {Pillai},
  \citenamefont {Arpanaei},\ and\ \citenamefont {Kingshott}}]{Singh2011}%
  \BibitemOpen
  \bibfield  {author} {\bibinfo {author} {\bibfnamefont {G.}~\bibnamefont
  {Singh}}, \bibinfo {author} {\bibfnamefont {S.}~\bibnamefont {Pillai}},
  \bibinfo {author} {\bibfnamefont {A.}~\bibnamefont {Arpanaei}},\ and\
  \bibinfo {author} {\bibfnamefont {P.}~\bibnamefont {Kingshott}},\ }\bibfield
  {title} {\enquote {\bibinfo {title} {Electrostatic and capillary force
  directed tunable 3{D} binary micro- and nanoparticle assemblies on
  surfaces},}\ }\href {https://doi.org/10.1088/0957-4484/22/22/225601}
  {\bibfield  {journal} {\bibinfo  {journal} {Nanotechnology}\ }\textbf
  {\bibinfo {volume} {22}},\ \bibinfo {pages} {225601} (\bibinfo {year}
  {2011}{\natexlab{a}})}\BibitemShut {NoStop}%
\bibitem [{\citenamefont {Singh}\ \emph
  {et~al.}(2011{\natexlab{b}})\citenamefont {Singh}, \citenamefont {Gohri},
  \citenamefont {Pillai}, \citenamefont {Arpanaei}, \citenamefont {Foss},\ and\
  \citenamefont {Kingshott}}]{Singh2011acsNano}%
  \BibitemOpen
  \bibfield  {author} {\bibinfo {author} {\bibfnamefont {G.}~\bibnamefont
  {Singh}}, \bibinfo {author} {\bibfnamefont {V.}~\bibnamefont {Gohri}},
  \bibinfo {author} {\bibfnamefont {S.}~\bibnamefont {Pillai}}, \bibinfo
  {author} {\bibfnamefont {A.}~\bibnamefont {Arpanaei}}, \bibinfo {author}
  {\bibfnamefont {M.}~\bibnamefont {Foss}},\ and\ \bibinfo {author}
  {\bibfnamefont {P.}~\bibnamefont {Kingshott}},\ }\bibfield  {title} {\enquote
  {\bibinfo {title} {Large-area protein patterns generated by ordered binary
  colloidal assemblies as templates},}\ }\href
  {https://doi.org/10.1021/nn102867z} {\bibfield  {journal} {\bibinfo
  {journal} {{ACS} Nano}\ }\textbf {\bibinfo {volume} {5}},\ \bibinfo {pages}
  {3542--3551} (\bibinfo {year} {2011}{\natexlab{b}})}\BibitemShut {NoStop}%
\bibitem [{\citenamefont {Singh}\ \emph
  {et~al.}(2011{\natexlab{c}})\citenamefont {Singh}, \citenamefont {Pillai},
  \citenamefont {Arpanaei},\ and\ \citenamefont
  {Kingshott}}]{Singh2011advMater}%
  \BibitemOpen
  \bibfield  {author} {\bibinfo {author} {\bibfnamefont {G.}~\bibnamefont
  {Singh}}, \bibinfo {author} {\bibfnamefont {S.}~\bibnamefont {Pillai}},
  \bibinfo {author} {\bibfnamefont {A.}~\bibnamefont {Arpanaei}},\ and\
  \bibinfo {author} {\bibfnamefont {P.}~\bibnamefont {Kingshott}},\ }\bibfield
  {title} {\enquote {\bibinfo {title} {Highly ordered mixed protein patterns
  over large areas from self-assembly of binary colloids},}\ }\href
  {https://doi.org/10.1002/adma.201004657} {\bibfield  {journal} {\bibinfo
  {journal} {Advanced Materials}\ }\textbf {\bibinfo {volume} {23}},\ \bibinfo
  {pages} {1519--1523} (\bibinfo {year} {2011}{\natexlab{c}})}\BibitemShut
  {NoStop}%
\bibitem [{\citenamefont {Singh}\ \emph
  {et~al.}(2011{\natexlab{d}})\citenamefont {Singh}, \citenamefont {Griesser},
  \citenamefont {Bremmell},\ and\ \citenamefont {Kingshott}}]{Singh2011adfm}%
  \BibitemOpen
  \bibfield  {author} {\bibinfo {author} {\bibfnamefont {G.}~\bibnamefont
  {Singh}}, \bibinfo {author} {\bibfnamefont {H.~J.}\ \bibnamefont {Griesser}},
  \bibinfo {author} {\bibfnamefont {K.}~\bibnamefont {Bremmell}},\ and\
  \bibinfo {author} {\bibfnamefont {P.}~\bibnamefont {Kingshott}},\ }\bibfield
  {title} {\enquote {\bibinfo {title} {Highly ordered nanometer-scale chemical
  and protein patterns by binary colloidal crystal lithography combined with
  plasma polymerization},}\ }\href {https://doi.org/10.1002/adfm.201001340}
  {\bibfield  {journal} {\bibinfo  {journal} {Advanced Functional Materials}\
  }\textbf {\bibinfo {volume} {21}},\ \bibinfo {pages} {540--546} (\bibinfo
  {year} {2011}{\natexlab{d}})}\BibitemShut {NoStop}%
\bibitem [{\citenamefont {Singh}\ \emph
  {et~al.}(2011{\natexlab{e}})\citenamefont {Singh}, \citenamefont {Pillai},
  \citenamefont {Arpanaei},\ and\ \citenamefont {Kingshott}}]{Singh2011adfm21}%
  \BibitemOpen
  \bibfield  {author} {\bibinfo {author} {\bibfnamefont {G.}~\bibnamefont
  {Singh}}, \bibinfo {author} {\bibfnamefont {S.}~\bibnamefont {Pillai}},
  \bibinfo {author} {\bibfnamefont {A.}~\bibnamefont {Arpanaei}},\ and\
  \bibinfo {author} {\bibfnamefont {P.}~\bibnamefont {Kingshott}},\ }\bibfield
  {title} {\enquote {\bibinfo {title} {Layer-by-layer growth of multicomponent
  colloidal crystals over large areas},}\ }\href
  {https://doi.org/10.1002/adfm.201002716} {\bibfield  {journal} {\bibinfo
  {journal} {Advanced Functional Materials}\ }\textbf {\bibinfo {volume}
  {21}},\ \bibinfo {pages} {2556--2563} (\bibinfo {year}
  {2011}{\natexlab{e}})}\BibitemShut {NoStop}%
\bibitem [{\citenamefont {Parsa}\ \emph {et~al.}(2017)\citenamefont {Parsa},
  \citenamefont {Harmand}, \citenamefont {Sefiane}, \citenamefont {Bigerelle},\
  and\ \citenamefont {Deltombe}}]{Parsa2017}%
  \BibitemOpen
  \bibfield  {author} {\bibinfo {author} {\bibfnamefont {M.}~\bibnamefont
  {Parsa}}, \bibinfo {author} {\bibfnamefont {S.}~\bibnamefont {Harmand}},
  \bibinfo {author} {\bibfnamefont {K.}~\bibnamefont {Sefiane}}, \bibinfo
  {author} {\bibfnamefont {M.}~\bibnamefont {Bigerelle}},\ and\ \bibinfo
  {author} {\bibfnamefont {R.}~\bibnamefont {Deltombe}},\ }\bibfield  {title}
  {\enquote {\bibinfo {title} {Effect of substrate temperature on pattern
  formation of bidispersed particles from volatile drops},}\ }\href
  {https://doi.org/10.1021/acs.jpcb.7b09700} {\bibfield  {journal} {\bibinfo
  {journal} {The Journal of Physical Chemistry B}\ }\textbf {\bibinfo {volume}
  {121}},\ \bibinfo {pages} {11002--11017} (\bibinfo {year}
  {2017})}\BibitemShut {NoStop}%
\bibitem [{\citenamefont {Patil}, \citenamefont {Bhardwaj},\ and\ \citenamefont
  {Sharma}(2018)}]{Patil2018}%
  \BibitemOpen
  \bibfield  {author} {\bibinfo {author} {\bibfnamefont {N.~D.}\ \bibnamefont
  {Patil}}, \bibinfo {author} {\bibfnamefont {R.}~\bibnamefont {Bhardwaj}},\
  and\ \bibinfo {author} {\bibfnamefont {A.}~\bibnamefont {Sharma}},\
  }\bibfield  {title} {\enquote {\bibinfo {title} {Self-sorting of bidispersed
  colloidal particles near contact line of an evaporating sessile droplet},}\
  }\href {https://doi.org/10.1021/acs.langmuir.8b00427} {\bibfield  {journal}
  {\bibinfo  {journal} {Langmuir}\ }\textbf {\bibinfo {volume} {34}},\ \bibinfo
  {pages} {12058--12070} (\bibinfo {year} {2018})}\BibitemShut {NoStop}%
\bibitem [{\citenamefont {Wong}\ \emph {et~al.}(2011)\citenamefont {Wong},
  \citenamefont {Chen}, \citenamefont {Shen},\ and\ \citenamefont
  {Ho}}]{Wong2011}%
  \BibitemOpen
  \bibfield  {author} {\bibinfo {author} {\bibfnamefont {T.-S.}\ \bibnamefont
  {Wong}}, \bibinfo {author} {\bibfnamefont {T.-H.}\ \bibnamefont {Chen}},
  \bibinfo {author} {\bibfnamefont {X.}~\bibnamefont {Shen}},\ and\ \bibinfo
  {author} {\bibfnamefont {C.-M.}\ \bibnamefont {Ho}},\ }\bibfield  {title}
  {\enquote {\bibinfo {title} {Nanochromatography driven by the coffee ring
  effect},}\ }\href {https://doi.org/10.1021/ac102963x} {\bibfield  {journal}
  {\bibinfo  {journal} {Analytical Chemistry}\ }\textbf {\bibinfo {volume}
  {83}},\ \bibinfo {pages} {1871--1873} (\bibinfo {year} {2011})}\BibitemShut
  {NoStop}%
\bibitem [{\citenamefont {Monteux}\ and\ \citenamefont
  {Lequeux}(2011)}]{Monteux2011}%
  \BibitemOpen
  \bibfield  {author} {\bibinfo {author} {\bibfnamefont {C.}~\bibnamefont
  {Monteux}}\ and\ \bibinfo {author} {\bibfnamefont {F.}~\bibnamefont
  {Lequeux}},\ }\bibfield  {title} {\enquote {\bibinfo {title} {Packing and
  sorting colloids at the contact line of a drying drop},}\ }\href
  {https://doi.org/10.1021/la104055j} {\bibfield  {journal} {\bibinfo
  {journal} {Langmuir}\ }\textbf {\bibinfo {volume} {27}},\ \bibinfo {pages}
  {2917--2922} (\bibinfo {year} {2011})}\BibitemShut {NoStop}%
\bibitem [{\citenamefont {Iqbal}\ \emph {et~al.}(2018)\citenamefont {Iqbal},
  \citenamefont {Majhy}, \citenamefont {Shen},\ and\ \citenamefont
  {Sen}}]{Iqbal2018}%
  \BibitemOpen
  \bibfield  {author} {\bibinfo {author} {\bibfnamefont {R.}~\bibnamefont
  {Iqbal}}, \bibinfo {author} {\bibfnamefont {B.}~\bibnamefont {Majhy}},
  \bibinfo {author} {\bibfnamefont {A.~Q.}\ \bibnamefont {Shen}},\ and\
  \bibinfo {author} {\bibfnamefont {A.~K.}\ \bibnamefont {Sen}},\ }\bibfield
  {title} {\enquote {\bibinfo {title} {Evaporation and morphological patterns
  of bi-dispersed colloidal droplets on hydrophilic and hydrophobic
  surfaces},}\ }\href {https://doi.org/10.1039/c8sm01915k} {\bibfield
  {journal} {\bibinfo  {journal} {Soft Matter}\ }\textbf {\bibinfo {volume}
  {14}},\ \bibinfo {pages} {9901--9909} (\bibinfo {year} {2018})}\BibitemShut
  {NoStop}%
\bibitem [{\citenamefont {Marinaro}, \citenamefont {Riekel},\ and\
  \citenamefont {Gentile}(2021)}]{Marinaro2021}%
  \BibitemOpen
  \bibfield  {author} {\bibinfo {author} {\bibfnamefont {G.}~\bibnamefont
  {Marinaro}}, \bibinfo {author} {\bibfnamefont {C.}~\bibnamefont {Riekel}},\
  and\ \bibinfo {author} {\bibfnamefont {F.}~\bibnamefont {Gentile}},\
  }\bibfield  {title} {\enquote {\bibinfo {title} {Size-exclusion particle
  separation driven by micro-flows in a quasi-spherical droplet: Modelling and
  experimental results},}\ }\href {https://doi.org/10.3390/mi12020185}
  {\bibfield  {journal} {\bibinfo  {journal} {Micromachines}\ }\textbf
  {\bibinfo {volume} {12}},\ \bibinfo {pages} {185} (\bibinfo {year}
  {2021})}\BibitemShut {NoStop}%
\bibitem [{\citenamefont {Upadhyay}\ and\ \citenamefont
  {Bhardwaj}(2021)}]{Upadhyay2021}%
  \BibitemOpen
  \bibfield  {author} {\bibinfo {author} {\bibfnamefont {G.}~\bibnamefont
  {Upadhyay}}\ and\ \bibinfo {author} {\bibfnamefont {R.}~\bibnamefont
  {Bhardwaj}},\ }\bibfield  {title} {\enquote {\bibinfo {title} {Colloidal
  deposits via capillary bridge evaporation and particle sorting thereof},}\
  }\href {https://doi.org/10.1021/acs.langmuir.1c01869} {\bibfield  {journal}
  {\bibinfo  {journal} {Langmuir}\ }\textbf {\bibinfo {volume} {37}},\ \bibinfo
  {pages} {12071--12088} (\bibinfo {year} {2021})}\BibitemShut {NoStop}%
\bibitem [{\citenamefont {Hu}\ \emph {et~al.}(2021)\citenamefont {Hu},
  \citenamefont {Wang}, \citenamefont {Yang},\ and\ \citenamefont
  {Chen}}]{Hu2021}%
  \BibitemOpen
  \bibfield  {author} {\bibinfo {author} {\bibfnamefont {T.-Y.}\ \bibnamefont
  {Hu}}, \bibinfo {author} {\bibfnamefont {C.}~\bibnamefont {Wang}}, \bibinfo
  {author} {\bibfnamefont {K.-C.}\ \bibnamefont {Yang}},\ and\ \bibinfo
  {author} {\bibfnamefont {L.-J.}\ \bibnamefont {Chen}},\ }\bibfield  {title}
  {\enquote {\bibinfo {title} {Gravity effect of silica and polystyrene
  particles on deposition pattern control and particle size distribution on
  hydrophobic surfaces},}\ }\href {https://doi.org/10.1016/j.jiec.2021.10.030}
  {\bibfield  {journal} {\bibinfo  {journal} {Journal of Industrial and
  Engineering Chemistry}\ } (\bibinfo {year} {2021}),\
  10.1016/j.jiec.2021.10.030}\BibitemShut {NoStop}%
\bibitem [{\citenamefont {Kolegov}\ and\ \citenamefont
  {Barash}(2019)}]{Kolegov2019}%
  \BibitemOpen
  \bibfield  {author} {\bibinfo {author} {\bibfnamefont {K.~S.}\ \bibnamefont
  {Kolegov}}\ and\ \bibinfo {author} {\bibfnamefont {L.~Y.}\ \bibnamefont
  {Barash}},\ }\bibfield  {title} {\enquote {\bibinfo {title} {Joint effect of
  advection, diffusion, and capillary attraction on the spatial structure of
  particle depositions from evaporating droplets},}\ }\href
  {https://doi.org/10.1103/physreve.100.033304} {\bibfield  {journal} {\bibinfo
   {journal} {Physical Review E}\ }\textbf {\bibinfo {volume} {100}} (\bibinfo
  {year} {2019}),\ 10.1103/physreve.100.033304}\BibitemShut {NoStop}%
\bibitem [{\citenamefont {Zolotarev}\ and\ \citenamefont
  {Kolegov}(2021)}]{Zolotarev2021}%
  \BibitemOpen
  \bibfield  {author} {\bibinfo {author} {\bibfnamefont {P.~A.}\ \bibnamefont
  {Zolotarev}}\ and\ \bibinfo {author} {\bibfnamefont {K.~S.}\ \bibnamefont
  {Kolegov}},\ }\bibfield  {title} {\enquote {\bibinfo {title} {Average cluster
  size inside sediment left after droplet desiccation},}\ }\href
  {https://doi.org/10.1088/1742-6596/1740/1/012029} {\bibfield  {journal}
  {\bibinfo  {journal} {Journal of Physics: Conference Series}\ }\textbf
  {\bibinfo {volume} {1740}},\ \bibinfo {pages} {012029} (\bibinfo {year}
  {2021})}\BibitemShut {NoStop}%
\bibitem [{\citenamefont {Larson}(2014)}]{Larson2014}%
  \BibitemOpen
  \bibfield  {author} {\bibinfo {author} {\bibfnamefont {R.~G.}\ \bibnamefont
  {Larson}},\ }\bibfield  {title} {\enquote {\bibinfo {title} {Transport and
  deposition patterns in drying sessile droplets},}\ }\href
  {https://doi.org/10.1002/aic.14338} {\bibfield  {journal} {\bibinfo
  {journal} {{AIChE} Journal}\ }\textbf {\bibinfo {volume} {60}},\ \bibinfo
  {pages} {1538--1571} (\bibinfo {year} {2014})}\BibitemShut {NoStop}%
\bibitem [{\citenamefont {Li}, \citenamefont {Fan},\ and\ \citenamefont
  {Yin}(2021)}]{Li2021}%
  \BibitemOpen
  \bibfield  {author} {\bibinfo {author} {\bibfnamefont {Z.}~\bibnamefont
  {Li}}, \bibinfo {author} {\bibfnamefont {Q.}~\bibnamefont {Fan}},\ and\
  \bibinfo {author} {\bibfnamefont {Y.}~\bibnamefont {Yin}},\ }\bibfield
  {title} {\enquote {\bibinfo {title} {Colloidal self-assembly approaches to
  smart nanostructured materials},}\ }\href
  {https://doi.org/10.1021/acs.chemrev.1c00482} {\bibfield  {journal} {\bibinfo
   {journal} {Chemical Reviews}\ } (\bibinfo {year} {2021}),\
  10.1021/acs.chemrev.1c00482}\BibitemShut {NoStop}%
\bibitem [{\citenamefont {Ortega}, \citenamefont {Ritacco},\ and\ \citenamefont
  {Rubio}(2010)}]{Ortega2010}%
  \BibitemOpen
  \bibfield  {author} {\bibinfo {author} {\bibfnamefont {F.}~\bibnamefont
  {Ortega}}, \bibinfo {author} {\bibfnamefont {H.}~\bibnamefont {Ritacco}},\
  and\ \bibinfo {author} {\bibfnamefont {R.~G.}\ \bibnamefont {Rubio}},\
  }\bibfield  {title} {\enquote {\bibinfo {title} {Interfacial microrheology:
  {P}article tracking and related techniques},}\ }\href
  {https://doi.org/10.1016/j.cocis.2010.03.001} {\bibfield  {journal} {\bibinfo
   {journal} {Current Opinion in Colloid {\&} Interface Science}\ }\textbf
  {\bibinfo {volume} {15}},\ \bibinfo {pages} {237--245} (\bibinfo {year}
  {2010})}\BibitemShut {NoStop}%
\bibitem [{\citenamefont {Deshmukh}\ \emph {et~al.}(2015)\citenamefont
  {Deshmukh}, \citenamefont {van~den Ende}, \citenamefont {Stuart},
  \citenamefont {Mugele},\ and\ \citenamefont {Duits}}]{Deshmukh2015}%
  \BibitemOpen
  \bibfield  {author} {\bibinfo {author} {\bibfnamefont {O.~S.}\ \bibnamefont
  {Deshmukh}}, \bibinfo {author} {\bibfnamefont {D.}~\bibnamefont {van~den
  Ende}}, \bibinfo {author} {\bibfnamefont {M.~C.}\ \bibnamefont {Stuart}},
  \bibinfo {author} {\bibfnamefont {F.}~\bibnamefont {Mugele}},\ and\ \bibinfo
  {author} {\bibfnamefont {M.~H.~G.}\ \bibnamefont {Duits}},\ }\bibfield
  {title} {\enquote {\bibinfo {title} {Hard and soft colloids at fluid
  interfaces: {A}dsorption, interactions, assembly {\&} rheology},}\ }\href
  {https://doi.org/10.1016/j.cis.2014.09.003} {\bibfield  {journal} {\bibinfo
  {journal} {Advances in Colloid and Interface Science}\ }\textbf {\bibinfo
  {volume} {222}},\ \bibinfo {pages} {215--227} (\bibinfo {year}
  {2015})}\BibitemShut {NoStop}%
\bibitem [{\citenamefont {Lotito}\ and\ \citenamefont
  {Zambelli}(2019)}]{Lotito2019}%
  \BibitemOpen
  \bibfield  {author} {\bibinfo {author} {\bibfnamefont {V.}~\bibnamefont
  {Lotito}}\ and\ \bibinfo {author} {\bibfnamefont {T.}~\bibnamefont
  {Zambelli}},\ }\bibfield  {title} {\enquote {\bibinfo {title} {A journey
  through the landscapes of small particles in binary colloidal assemblies:
  {U}nveiling structural transitions from isolated particles to clusters upon
  variation in composition},}\ }\href {https://doi.org/10.3390/nano9070921}
  {\bibfield  {journal} {\bibinfo  {journal} {Nanomaterials}\ }\textbf
  {\bibinfo {volume} {9}},\ \bibinfo {pages} {921} (\bibinfo {year}
  {2019})}\BibitemShut {NoStop}%
\bibitem [{\citenamefont {Lotito}\ and\ \citenamefont
  {Zambelli}(2020)}]{Lotito2020}%
  \BibitemOpen
  \bibfield  {author} {\bibinfo {author} {\bibfnamefont {V.}~\bibnamefont
  {Lotito}}\ and\ \bibinfo {author} {\bibfnamefont {T.}~\bibnamefont
  {Zambelli}},\ }\bibfield  {title} {\enquote {\bibinfo {title} {Pattern
  detection in colloidal assembly: {A} mosaic of analysis techniques},}\ }\href
  {https://doi.org/10.1016/j.cis.2020.102252} {\bibfield  {journal} {\bibinfo
  {journal} {Advances in Colloid and Interface Science}\ }\textbf {\bibinfo
  {volume} {284}},\ \bibinfo {pages} {102252} (\bibinfo {year}
  {2020})}\BibitemShut {NoStop}%
\bibitem [{\citenamefont {Jung}, \citenamefont {Kim},\ and\ \citenamefont
  {Yoo}(2009)}]{Jung2009}%
  \BibitemOpen
  \bibfield  {author} {\bibinfo {author} {\bibfnamefont {J.-Y.}\ \bibnamefont
  {Jung}}, \bibinfo {author} {\bibfnamefont {Y.~W.}\ \bibnamefont {Kim}},\ and\
  \bibinfo {author} {\bibfnamefont {J.~Y.}\ \bibnamefont {Yoo}},\ }\bibfield
  {title} {\enquote {\bibinfo {title} {Behavior of particles in an evaporating
  didisperse colloid droplet on a hydrophilic surface},}\ }\href
  {https://doi.org/10.1021/ac901247c} {\bibfield  {journal} {\bibinfo
  {journal} {Analytical Chemistry}\ }\textbf {\bibinfo {volume} {81}},\
  \bibinfo {pages} {8256--8259} (\bibinfo {year} {2009})}\BibitemShut {NoStop}%
\bibitem [{\citenamefont {yeul Jung}\ \emph {et~al.}(2010)\citenamefont {yeul
  Jung}, \citenamefont {Kim}, \citenamefont {Yoo}, \citenamefont {Koo},\ and\
  \citenamefont {Kang}}]{Jung2010}%
  \BibitemOpen
  \bibfield  {author} {\bibinfo {author} {\bibfnamefont {J.}~\bibnamefont {yeul
  Jung}}, \bibinfo {author} {\bibfnamefont {Y.~W.}\ \bibnamefont {Kim}},
  \bibinfo {author} {\bibfnamefont {J.~Y.}\ \bibnamefont {Yoo}}, \bibinfo
  {author} {\bibfnamefont {J.}~\bibnamefont {Koo}},\ and\ \bibinfo {author}
  {\bibfnamefont {Y.~T.}\ \bibnamefont {Kang}},\ }\bibfield  {title} {\enquote
  {\bibinfo {title} {Forces acting on a single particle in an evaporating
  sessile droplet on a hydrophilic surface},}\ }\href
  {https://doi.org/10.1021/ac902288z} {\bibfield  {journal} {\bibinfo
  {journal} {Analytical Chemistry}\ }\textbf {\bibinfo {volume} {82}},\
  \bibinfo {pages} {784--788} (\bibinfo {year} {2010})}\BibitemShut {NoStop}%
\bibitem [{\citenamefont {Lebedev-Stepanov}\ and\ \citenamefont
  {Vlasov}(2013)}]{LebedevStepanov2013}%
  \BibitemOpen
  \bibfield  {author} {\bibinfo {author} {\bibfnamefont {P.}~\bibnamefont
  {Lebedev-Stepanov}}\ and\ \bibinfo {author} {\bibfnamefont {K.}~\bibnamefont
  {Vlasov}},\ }\bibfield  {title} {\enquote {\bibinfo {title} {Simulation of
  self-assembly in an evaporating droplet of colloidal solution by dissipative
  particle dynamics},}\ }\href {https://doi.org/10.1016/j.colsurfa.2013.05.012}
  {\bibfield  {journal} {\bibinfo  {journal} {Colloids and Surfaces A:
  Physicochemical and Engineering Aspects}\ }\textbf {\bibinfo {volume}
  {432}},\ \bibinfo {pages} {132--138} (\bibinfo {year} {2013})}\BibitemShut
  {NoStop}%
\end{thebibliography}%

\end{document}